\documentclass[notitlepage, onecolumn, floatfix,nofootinbib,british]{revtex4-1}
\usepackage{graphicx}
\usepackage{amsmath,mathrsfs}
\usepackage{amssymb}
\PassOptionsToPackage{svgnames}{xcolor}
\usepackage[usenames,dvipsnames]{xcolor}
\usepackage{siunitx}
\usepackage{bbm}
\usepackage{hyperref}
\usepackage[utf8]{inputenc}
\usepackage{calc}
\usepackage{fancyhdr}
\usepackage[T1]{fontenc}
\usepackage{fancybox}
\usepackage{textcomp}
\usepackage{float}
\usepackage{dcolumn}
\usepackage{tikz}
\usepackage{rotating}
\usepackage{multirow}
\usepackage{booktabs}
\usepackage{slashed}
\usepackage{color}
\usepackage[capitalise]{cleveref}

\hypersetup{colorlinks=true
,urlcolor=blue
,anchorcolor=blue
,citecolor=blue
,filecolor=blue
,linkcolor=blue
,menucolor=blue
,pagecolor=blue
,linktocpage=true
,pdfproducer=medialab
,pdfa=true
}
\newcommand{\qR[1]}{R_{#1}}
\newcommand{\Q}{Q}
\newcommand{\Ld}{L}
\newcommand{\Hu}{H_u}
\newcommand{\Hd}{H_d}
\newcommand{\D}{\bar{D}}
\newcommand{\U}{\bar{U}}
\newcommand{\E}{\bar{E}}
\newcommand{\N}{\bar{N}}
\newcommand{\ZR}{\mathbb{Z}_4^R}
\newcommand{\mgrav}{M_{\text{grav}}}

\newcommand{\dd}{\Delta_{21}^L}
\newcommand{\DD}{\Delta_{31}^L}
\newcommand{\DDij}{\Delta_{ij}^L}
\newcommand{\Ws}{\,\mathcal{W}}
\newcommand{\Cs}{\,\mathcal{C}}
\newcommand{\dsi}{\,\mathcal{N}_d}
\newcommand{\dsc}{C_4}
\newcommand{\kosp}{S}
\newcommand{\Dh}{\Delta^H}
\newcommand{\squi}{\zeta}
\newcommand{\UR}{\text U(1)_R}
\newcommand{\wt}{\widetilde}
\newcommand{\be}{\begin{equation}}
\newcommand{\ee}{\end{equation}}
\newcommand{\bea}{\begin{eqnarray}}
\newcommand{\eea}{\end{eqnarray}}

\makeatletter
\providecommand*{\diff}%
	{\@ifnextchar^{\DIfF}{\DIfF^{}}}
\def\DIfF^#1{%
	\mathop{\mathrm{\mathstrut d}}%
		\nolimits^{#1}\gobblespace}
\def\gobblespace{%
	\futurelet\diffarg\opspace}
\def\opspace{%
	\let\DiffSpace\!%
	\ifx\diffarg(%
		\let\DiffSpace\relax
	\else
		\ifx\diffarg[%
			\let\DiffSpace\relax
		\else
  			\ifx\diffarg\{%
				\let\DiffSpace\relax
			\fi\fi\fi\DiffSpace}

\newcommand{\Rmnum}[1]{\expandafter\@slowromancap\romannumeral #1@}
\newcommand{\GeV}{{GeV}}
\newcommand{\TeV}{{TeV}}
\newcommand{\eV}{{eV}}

\newcolumntype{C}{>{\centering\arraybackslash$}p{0.85cm}<{$}}

\begin{document}

\title{Froggatt-Nielsen models with a residual $\boldsymbol{\ZR}$ symmetry}%

\author{Herbi K. Dreiner}%
\email{dreiner@th.physik.uni-bonn.de}
\author{Toby Opferkuch}%
\email{toby@th.physik.uni-bonn.de}
\affiliation{Physikalisches Institut and 
Bethe Center for Theoretical Physics, 
University of Bonn, 53115 Bonn, Germany}

\author{Christoph Luhn}%
\email{christoph.luhn@durham.ac.uk}
\affiliation{Institute for Particle Physics Phenomenology, University of Durham,
Durham, DH1 3LE, United Kingdom}

\begin{abstract}
  The Froggatt-Nielsen mechanism provides an elegant explanation for
  the hierarchies of fermion masses and mixings in terms of a $\text
  U(1)$ symmetry.  Promoting such a family symmetry to an
  $R$-symmetry, we explicitly construct supersymmetric
    Froggatt-Nielsen models which are gauged, family dependent $\UR$
    completions of the $\ZR$ symmetry proposed by Lee, Raby, Ratz,
    Ross, Schieren, Schmidt-Hoberg and Vaudrevange in 2010. Forbidden by
    $\ZR$, the $\mu$-term is generated around the supersymmetry
    breaking scale $m_{3/2}$ from either the K\"ahler potential or the
    superpotential.  Neutrinos acquire their mass via the type~I
    seesaw mechanism with three right-handed neutrino
    superfields. Taking into account the Green-Schwarz anomaly
    cancellation conditions, we arrive at a total of 3~$\times$~34 distinct
    phenomenologically viable charge assignments for the standard
    model fields, most of which feature highly fractional charges.
\end{abstract}

\maketitle

\section{Introduction}

The standard model of particle physics postulates three families of
fermions. The masses and mixings of the quarks and leptons are dictated by
Yukawa couplings,
\be
Y^{ij} ~ \overline \psi_i \, \psi_j \,H \ ,
\ee
that is, interactions of two fermionic fields~$\overline\psi_i$
and~$\psi_j$ with the recently discovered Higgs
boson~$H$~\cite{Aad:2012tfa,Chatrchyan:2012ufa}. In order to correctly
describe the hierarchies observed in the fermionic masses and mixings
without relying on fine-tuned correlations among the entries of a
Yukawa matrix~$Y$, it is necessary to assume that~$Y$ itself features
a hierarchical pattern. Such a special structure of a priori random
coupling constants calls for an explanation. In 1979 one was proposed by
Froggatt and Nielsen in the form of an underlying family dependent  $\text U(1)$
gauge symmetry extension~\cite{Froggatt:1978nt}. At the effective
level, the Froggatt-Nielsen (\textsc{FN}) mechanism requires the
introduction of a standard model neutral but $\text U(1)$ charged
field which, like the Higgs, develops a vacuum expectation value
(\textsc{VEV}). Appropriate powers of this so-called flavon
field~$\phi$ compensate non-zero charges of the original trilinear
Yukawa terms, thereby giving rise to a hierarchy in the entries of the
Yukawa matrices. Despite motivating a hierarchical structure, the
exact value of each Yukawa interaction depends on unknown order-one
coefficients. The number of free parameters can only be reduced
compared to a theory which does not invoke the \textsc{FN} mechanism,
if the extra $\text U(1)$ family symmetry is spontaneously
broken down to a residual discrete $ \mathbb{Z}_ N$ symmetry.

Such a possibility has been proposed and studied carefully in
supersymmetric \textsc{FN} models with residual matter (or
equivalently $R$-) parity~\cite{Dreiner:2003yr}, baryon
triality~\cite{Dreiner:2006xw}, and proton
hexality~\cite{Dreiner:2007vp}, and it has been demonstrated how these
residual discrete symmetries are obtained exactly from certain simple
linear conditions on the $\text U(1)$ charge assignments. In the
present paper, we consider a similar situation where a discrete
$R$-symmetry ($\mathbb Z_N^R$) arises from a family dependent
gauged $\text U(1)_R$ symmetry. In particular, we focus on the
$\mathbb Z_4^R$ symmetry proposed in~\cite{Lee:2010gv,Lee:2011dya},
see also~\cite{Kurosawa:2001iq,Babu:2002tx}. This family independent
symmetry is compatible with grand unification and anomaly-free by
means of the (discrete) Green-Schwarz (\textsc{GS})
mechanism~\cite{Green:1984sg}. Furthermore, it provides an attractive
solution to the $\mu$-problem~\cite{Kim:1983dt} by forbidding the
bilinear Higgs superpotential term. The $\mu$-term
is generated dynamically at the electroweak scale when the
$\mathbb Z_4^R$ symmetry is broken to standard $R$-parity, $\mathbb
Z_2^R$,\footnote{Here, $R$-parity arises from an $R$-symmetry. Note, however,
that it can equally well occur in the form of matter parity, a non-$R$
symmetry which is equivalent to $R$-parity.} by the breakdown of supersymmetry,
either in the K\"ahler potential~\cite{Giudice:1988yz} or in the
superpotential~\cite{Kim:1994eu}. Starting with a $\text U(1)_R$
family symmetry, the class of Froggatt-Nielsen models constructed in
this paper is characterized by the following symmetry breaking chain,
\be
\label{eq:breakingprocedure}
\textsc{SM} \times \text{U}(1)_R 
~~~\xrightarrow{\begin{array}{c}{\mathrm{family~symmetry}}\\\mathrm{breaking}\end{array}} ~~~
\textsc{SM} \times \ZR 
~~~\xrightarrow{\begin{array}{c}{\mathrm{supersymmetry}}\\\mathrm{breaking}\end{array}}
~~~
\textsc{SM} \times \mathbb{Z}^R_2 \ ,
\ee
where \textsc{SM} refers to the standard model gauge group
$\text{SU}(3)_C\times \text{SU}(2)_W \times \text{U}(1)_Y$.  Our 
main concern shall be the first step, where we identify constraints on
the $\text U(1)_R$ charge assignments originating from the following
requirements:
\begin{itemize}
\item phenomenologically acceptable quark and lepton masses and
  mixings, including the neutrinos,
\item absence of the $\mu$-term,
\item anomaly freedom of the gauged $\text U(1)_R$ family symmetry via  the
  Green-Schwarz mechanism, 
\item a flavon \textsc{VEV}  $\langle \phi \rangle \sim \epsilon M_{\mathrm{grav}}$
  induced by the Dine-Seiberg-Wen-Witten
  mechanism~\cite{Dine:1986zy,Dine:1987bq,Atick:1987gy,Dine:1987gj} (here 
 $\epsilon \sim 0.2$ is an expansion parameter similar in size to the
  Wolfenstein parameter $\lambda_c$ of the Cabibbo-Kobayashi-Maskawa (\textsc{CKM})
  matrix, and $M_{\mathrm{grav}} \sim  \SI{2.4e18}{\GeV}$ denotes the gravitational
  scale),
  \item existence of a residual $\ZR$ symmetry after $\text U(1)_R$ breaking.
\end{itemize}

The paper is organized as follows. In \cref{sec:R-FN} we discuss how
the hierarchies of the fermion masses and mixings can arise from a
family dependent $\text U(1)_R$ symmetry \`a~la Froggatt and
Nielsen. In \cref{sec:anomalies} we formulate the anomaly constraints
on possible $\text U(1)_R$ charge assignments, and show how
non-vanishing anomaly coefficients fix the \textsc{VEV} of the flavon
field. Demanding an exact residual $\ZR$ symmetry, we derive the
corresponding linear conditions in \cref{sec:z4r}.  The neutrino
constraints are studied in \cref{sec:neutrino}. Collecting all
results, we arrive at a set of 3~$\times$~34 possible charge assignments which
are listed in \cref{app:tables}. We then conclude in \cref{sec:conclusion}.

\section{Froggatt-Nielsen mechanism using a $\boldsymbol{\text U(1)_R}$ symmetry}
\label{sec:R-FN}

In this section we focus on the first step of the breaking chain
depicted in \cref{eq:breakingprocedure}. The Froggatt-Nielsen
mechanism~\cite{Froggatt:1978nt} provides a symmetry 
argument to explain the hierarchies of the fermion masses and mixings based on
a family dependent $\text U(1)$ extension of the \textsc{SM} gauge group.  We
examine the situation where the \textsc{FN} $\text U(1)$ family symmetry is
altered to a gauged U(1) $R$-symmetry~\cite{Freedman:1976uk}. The \textsc{VEV}
of the \textsc{SM} singlet flavon breaks the gauged $\text U(1)_R$ symmetry
down to $\ZR$. We defer the discussion of
anomalies~\cite{Chamseddine:1995gb,Castano:1995ci} as well as residual 
discrete symmetries to later sections, and consider first the changes
to the \textsc{FN} mechanism due to a non-trivially $R$-charged
superspace coordinate~$\theta$.

It is convenient to fix the Froggatt-Nielsen $\text U(1)_R$ charge
normalization such that the flavon field $\phi$ carries $R$-charge
$\qR[\phi]=-1$. Then the charge of the superspace variable $\theta$
takes the general value~$\qR[\theta]$, and a local $\text U(1)_R$
transformation, parametrized by $\alpha(x)$, maps 
\bea
\theta &~\rightarrow~& ~~~~\,~\theta' = e^{i \qR[\theta] \alpha(x)}\,\theta\ ,\\
 \int \diff \theta&~ \rightarrow ~& \int \diff \theta'= e^{-i \qR[\theta]
   \alpha(x)}\int \diff \theta\ .
\eea
As a consequence, a term allowed in the $\text U(1)_R$ symmetric
superpotential has to carry a charge of $+2\qR[\theta]$. The total charge
$\qR[\text{total}]$ of a term {\it with the flavon field removed} has to equal
$2\qR[\theta]+n$, where $n$ is a non-negative integer, in order 
to be generated effectively after $\text U(1)_R$ breaking. In the unbroken
phase, non-zero $n$ can be compensated by $n$ powers of the flavon field
which, in our normalization has $\text U(1)_R$ charge $\qR[\phi]=-1$. 

To illustrate the origin of hierarchical Yukawa matrices $Y$, let us
consider the example of the operator $Y_u^{ij} \Q_i \Hu
\U_j$. Denoting the total $\text U(1)_R$ charge of this operator by
$\qR[\text{total}]^{ij}=\qR[\Q_i]+\qR[\Hu]+\qR[\U_j]$,  it originates from an
underlying $\text U(1)_R$ symmetric term of the form
\begin{align}
y_u^{ij} \left(\frac{\phi}{\mgrav}\right)^{\qR[\text{total}]^{ij}-2\qR[\theta]}
\Theta\left[\qR[\text{total}]^{ij}-2\qR[\theta]\right]
~ \Q_i \Hu \U_j\ ,\label{eq:yuk-exampla}
\end{align}
where $\frac{1}{\sqrt{10}} \leq y_u^{ij} \leq \sqrt{10}$ are undetermined
order-one coefficients, $M_{\text{grav}}$ denotes the gravitational scale, and 
\begin{align}
\Theta\left[n\right] &= \left\{
 \begin{array}{l l}
1 & \quad n \in \mathbb{N}, \\
0 &\quad \text{otherwise},
\end{array}\right.
\end{align}
ensures that the superpotential remains holomorphic and satisfies the cluster
decomposition principle~\cite{CDP,Weinberg:1996kw}. When $\text U(1)_R$ is
broken by the flavon \textsc{VEV} $\langle \phi \rangle$, the hierarchies of the
Yukawa couplings $Y_u^{ij}$ are generated from \cref{eq:yuk-exampla}
due to family dependent and non-negative integer powers of the
expansion parameter $ \epsilon \equiv \frac{\langle \phi
  \rangle}{M_{\text{grav}}}< 1$.  In general, the \textsc{FN}
suppression of an operator with a total $\text U(1)_R$ charge of
$\qR[\text{total}]$ is given by $\epsilon^n$, with
$n={\qR[\text{total}]-2\qR[\theta]}$.  As we argue in \cref{sec:anomalies}, it
is possible to identify the expansion parameter $\epsilon$ with the
Wolfenstein parameter $\lambda_c\sim 0.22$ \cite{Wolfenstein:1983yz}.

The first goal is to formulate constraints on the $\UR$
charges such that the desired hierarchies of the GUT scale fermion
masses and mixings are
achieved~\cite{Ramond:1993kv,Binetruy:1994ru,Dudas:1995yu,Nir:1995bu,Binetruy:1996xk,Irges:1998ax,Jack:2002pn,Dreiner:2003hw,Dreiner:2003yr}. 
The singular values of the mass matrices \cite{Dreiner:2008tw} must be
related as
\begin{subequations}
\begin{align}
m_d:m_s:m_b &~\sim~ \lambda_c^4 :\lambda_c^2:1, \label{eq:dsb-ratios}\\
m_u : m_c :m_t &~\sim~ \lambda_c^8: \lambda_c^4 :1,\label{eq:uct-ratios}\\
m_e : m_\mu : m_\tau &~\sim~ \lambda_c^{4+z} : \lambda_c^2 :1 \label{eq:emutau-ratio},\\
m_b: m_t &~\sim ~\lambda_c^x \cot \beta, \label{eq:top-bottom-relation}\\
m_\tau : m_b &~\sim ~1,\label{eq:tau-bottom-relation}
\end{align}
\end{subequations}
where $z=0, 1$, $x=0,1,2,3$ and $\tan \beta = \frac{v_u}{v_d}$ is the ratio of
the up-type and the down-type Higgs \textsc{VEV}s. Furthermore, $m_t \sim v_u$,
and the \textsc{CKM} mixing matrix is parametrised by,
\begin{align} \label{eq:CKMmatrix}
 U_{\text{CKM}}\sim\left( \begin{array}{ccc}
1 & \lambda_c^{1+y}  & \lambda_c^{3+y} \\
\lambda_c^{1+y} & 1 & \lambda_c^2 \\
\lambda_c^{3+y} & \lambda_c^2 & 1 \end{array} \right),
\end{align}
where $y=-1,0,1$, with $y=0$ being the preferred value.
We note that the parametrisation in terms of $x,y,z$ allows for a high degree
of flexibility in which to interpret our final results for the $\text{U}(1)_R$
charge assignments.

In order to derive the physical masses and mixings in Froggatt-Nielsen models,
it is necessary to bring the kinetic terms into their canonical
form~\cite{Binetruy:1996xk,Dreiner:2003hw,Jack:2003pb,Dreiner:2003yr,Dreiner:2006xw}. Luckily,
the required non-unitary transformations do not change the \textsc{FN}
structure of the Yukawa couplings, but only affect the order-one coefficients which are
unknown anyway. Therefore, the left-chiral unitary transformations $V_{U_L}$ and
$V_{D_L}$ which are subsequently required to diagonalise the up-type and the
down-type quark mass matrices take the form
\be
V_{U_L}^{ij} \sim V_{D_L}^{ij} \sim
\epsilon^{|\qR[\Q_i]-\qR[\Q_j]|}\,,
\ee
leading to a CKM matrix of the same structure~\cite{Leurer:1992wg},
\be
U^{ij}_{\text{CKM}} ~=~ \left(V^{}_{U_L} V_{D_L}^\dagger\right)_{ij} \sim ~\epsilon^{|\qR[\Q_i]-\qR[\Q_j]|} \ .\label{eq:CKMbasis}
\ee
Enforcing the desired structure of the \textsc{CKM} matrix in
\cref{eq:CKMmatrix} and making use of the fact that successive
generations of quarks have smaller $\text{U}(1)_R$ charges,
i.e. $\qR[\Q_3]\leq \qR[\Q_2] \leq \qR[\Q_1]$, we can re-express 
$\qR[\Q_3]$ and $\qR[\Q_2]$ in terms of $\qR[\Q_1]$ and the parameter $y$,
\begin{align} \label{eq:Qgenrelation}
\qR[\Q_2]&=\qR[\Q_1]-1-y\ , \quad ~ \quad \qR[\Q_3]= \qR[\Q_1]-3-y\ . 
\end{align}
With these relations, dictated by the structure of the CKM matrix, the mass
ratios of \cref{eq:dsb-ratios,eq:uct-ratios,eq:emutau-ratio}
are realised if the $\UR$ charge assignments satisfy \cite{Dreiner:2007vp}
\begin{subequations}
\begin{align}
\label{eq:Dgenrelation}
\hspace{35mm}
\qR[\D_2]&=\qR[\D_1]-1+y\ ,  & \qR[\D_3]&=\qR[\D_1]-1+y\ , 
\hspace{35mm} \\
\label{eq:Ugenrelation}
\qR[\U_2]&=\qR[\U_1]-3+y\ ,  &\qR[\U_3]&=\qR[\U_1]-5+y \ ,\\
\label{eq:Egenrelation}
\qR[\E_2]&=\qR[\E_1]-2-z-\dd\ ,  & \qR[\E_3]&=\qR[\E_1]-4-z-\DD \ ,
\end{align}
\end{subequations}
where\footnote{$\DDij$ must necessarily be an integer as replacing $L_i$
by $L_{i'}$ ($i\neq i'$) in either a K\"ahler potential or superpotential term
must result in an allowed operator.} $\DDij=\qR[\Ld_i]-\qR[\Ld_j]$.
 Finally we can make use of
 \cref{eq:top-bottom-relation,eq:tau-bottom-relation} as well as
 the relation $m_t  \sim v_u$ to re-express the charges of all right-chiral
 superfields in terms of the charges of the left-chiral superfields. 
Inserting the relations of \cref{eq:Qgenrelation,eq:Dgenrelation,eq:Ugenrelation,eq:Egenrelation}, we readily find the conditions
\begin{subequations}
\begin{align}
\qR[\D_1]&= -\qR[\Q_1]-\qR[\Hd]+4+x+2\qR[\theta]\ ,\label{eq:d-rel}\\
\qR[\E_1]&= -\qR[\Ld_1]-\qR[\Hd]+4+x+z+2\qR[\theta]\ ,\label{eq:e-rel} \\
\qR[\U_1]&=-\qR[\Q_1]-\qR[\Hu]+8+2\qR[\theta]\label{eq:u-rel} \ .
\end{align}
\end{subequations}
Through these manipulations we have now re-expressed all charges in terms of
the integer parameters $x$, $y$, $z$, $\dd$ and $\DD$ as well as the
remaining charges $\qR[\Q_1]$, $\qR[\Ld_1]$, $\qR[\Hu]$, $\qR[\Hd]$ and,
since we are discussing $\UR$ symmetries, the charge $\qR[\theta]$ of the
superspace variable. 

In order to further reduce the number of free parameters we
impose anomaly freedom of the theory in the next section. However,  we first
comment on the possible origin of the $\mu$-term.  In principle, a term of the
form
\be
M  \left(\frac{\langle\phi\rangle}{M_{\text{grav}}}\right)^{|n|} \Hu \Hd \ ,\label{eq:muter}
\ee
can arise (effectively) in the superpotential either before or after
supersymmetry breaking, with the exponent $n$ given by
\be
n ~=~ \left\{
\begin{array}{ll}
n_b \equiv \qR[\Hu]+ \qR[\Hd]-2\qR[\theta] \ ,~~ & (\text{before SUSY breaking}) \ ,\\[2mm]
n_a \equiv \qR[\Hu]+ \qR[\Hd] \ ,\qquad & (\text{after SUSY breaking}) \ .
\end{array}
 \right.\label{eq:mu-n}
\ee
The former case is analogous to the generation of the Yukawa operators
discussed above, with $M=M_{\text{grav}}$, and requires the exponent $n=n_b$ to
take a non-negative integer value. To arrive at a $\mu$-parameter
around the order of the electroweak scale requires
$n=n_b>20$~\cite{Dreiner:2003yr}. Here we shall not pursue this direction
but rather consider the second option where the term in \cref{eq:muter} is
generated when supersymmetry is broken. This can happen in two
different ways~\cite{Brummer:2010fr,Lee:2010gv,Lee:2011dya}, both of
which require $n=n_a$ to take an {\it integer} value.

Adopting the Giudice-Masiero mechanism~\cite{Giudice:1988yz}, the
$\mu$-term arises from a non-renormalisable, non-holomorphic operator
in the K\"ahler potential, involving the complex conjugate of a
SM~$\times~\UR$ neutral hidden sector chiral superfield $Z$ whose
$F$-term acquires the \textsc{VEV} $\langle F_Z \rangle\sim
m_{3/2}M_{\text{grav}}$. This \textsc{VEV} breaks both supersymmetry
and the residual $\ZR$ symmetry, thereby effectively generating the
superpotential term in \cref{eq:muter} with $M$ being of the order of
the gravitino mass $m_{3/2}$. We emphasise that the Giudice-Masiero
mechanism is compatible with the parameter $n=n_a$ in \cref{eq:mu-n} being
either positive or negative, where the latter case requires one to use
the complex conjugate of the flavon field in \cref{eq:muter}.\footnote{In
Ref.~\cite{Giudice:1988yz} a global  U(1)$_R$ symmetry is invoked, whereas we
employ a local  one.}

Provided that the integer $n=n_a$ in \cref{eq:mu-n} is non-negative, an
effective $\mu$-term can additionally be obtained from a holomorphic
non-renormalisable superpotential operator \`a la Kim and
Nilles~\cite{Kim:1994eu}. In this case, the product $\left(\frac
{\phi}{M_{\text{grav}}}\right)^n \Hu \Hd$ is gauge neutral and can be
multiplied by $\frac{W_{\text {hidden}}}{M_{\text{grav}}^2}$ where
$W_{\text{hidden}}$ denotes the superpotential of a hidden
sector~\cite{Brummer:2010fr,Chen:2012jg}. When supersymmetry is
broken, the hidden sector superpotential develops a \textsc{VEV}
$\langle W_{\text{hidden}} \rangle \sim m_{3/2} M_{\text{grav}}^2$,
which in turn induces the effective superpotential term of
\cref{eq:muter} with $M\sim m_{3/2}$.  A possible realisation of this
mechanism is through a gaugino
condensate~\cite{Ferrara:1982qs,Derendinger:1985kk,Dine:1985rz}.

For the purpose of this paper, it is irrelevant whether we rely on the
Giudice-Masiero or the Kim-Nilles mechanism to generate the $\mu$-term
as both mechanisms result in $M\sim m_{3/2}$, see
\cref{eq:muter}. Moreover, both options require $n=n_a=\qR[\Hu]+\qR[\Hd]$ to
be an integer, which must be non-negative for the Kim-Nilles mechanism.
It is interesting to compare the above situation with the
conventional, i.e. non-$R$ \textsc{FN} scenario. For the conventional
case, in order to generate an effective $\mu$-term, the $\text U(1)$
charges of the two Higgs fields must add up to an integer $n^{}_{\qR
  [\theta]=0}$. For non-negative~$n^{}_{\qR[\theta]=0}$,  the
regular holomorphic contribution to the~$\mu$-term dominates
over the non-holomorphic Giudice-Masiero contribution. In order to
fall in the regime of an electroweak-scale~$\mu$-parameter,
$n^{}_{\qR[\theta]=0}>20$ is required. On the other hand, with
  negative~$n^{}_{\qR[\theta]=0}$, the holomorphic contribution is removed,
  and the Giudice-Masiero contribution leads to an attractive solution to the
  $\mu$-problem. 

In contrast, adopting a $\UR$ Froggatt-Nielsen symmetry allows to
solve the $\mu$-problem using either non-holomorphic terms \`a la
Giudice and Masiero or holomorphic terms \`a la Kim and Nilles. In
both cases, the $\UR$ charges of the two Higgs fields must add up to
an integer $n=n_a=\qR[\Hu]+\qR[\Hd]$. The regular superpotential term
(including powers of the flavon field $\phi$) which would typically dominate the
$\mu$-term, is forbidden as long as $\qR[\Hu]+\qR[\Hd]-2\qR
[\theta]\not\in \mathbb{N}$.

\section{Anomaly constraints}
\label{sec:anomalies}

To ensure a consistent theory after extending the \textsc{SM} gauge
group by a local \textsc{FN} family symmetry, one must enforce that
all anomaly coefficients are zero or that anomaly cancellation
proceeds via the Green-Schwarz mechanism~\cite{Green:1984sg}. Here we
make use of the latter option, or, more precisely its
four-dimensional analogue~\cite{Ibanez:1992fy}, as applied for
instance
in~\cite{Binetruy:1994ru,Dudas:1995yu,Nir:1995bu,Binetruy:1996xk,Irges:1998ax,Dreiner:2003hw,Dreiner:2003yr}.
We adopt a gauged $\UR$ symmetry, which was first discussed
in~\cite{Freedman:1976uk}. The anomalies were first studied
in~\cite{Chamseddine:1995gb,Castano:1995ci}. The components of a 
general left-chiral superfield $S=\varphi+\theta \psi +\theta^2 F$ transform as
\begin{align}
\varphi \rightarrow e^{i \qR[S] \alpha(x)}\varphi , \quad \quad \psi \rightarrow e^{i\left(\qR[S]-\qR[\theta]\right) \alpha(x)} \psi, \quad \quad F \rightarrow e^{i\left( \qR[S] -2 \qR[\theta]\right) \alpha(x)} F,
\end{align}
where $\qR[S]$ is the $\UR$ charge of $S$. The shifted charge $\qR[S]-\qR[\theta]$ of
the spin one-half particle $\psi$ contributes to the anomaly coefficients.
The gaugino component $\lambda$ of a gauge vector superfield~$V$ also
contributes to the anomaly coefficients. With $V$ being necessarily $\UR$
neutral, the gaugino carries charge $\qR[\theta]$ and transform as 
\begin{align}
\lambda \rightarrow e^{i \qR[\theta] \alpha(x)} \lambda.
\end{align}
With these remarks, the mixed $\text{SU}(3)_C-\text{SU}(3)_C-\text{U}(1)_R$
anomaly can be calculated easily, yielding~\cite{Dreiner:2012ae}
\begin{align}
A_{\text{SU}(3)_C-\text{SU}(3)_C-\text{U}(1)_R}&= \left(
\sum_{i=\text{coloured}}
\ell(\mathbf{r}_i)(\qR[S_i]-\qR[\theta])\right)+\overbrace{\ell(\mathbf{8})
  \qR[\theta]}^{\text{gluinos}}\ \notag \\
&=  \frac{1}{2} \sum_i \left(2 \qR[\Q_i]+\qR[\U_i]+\qR[\D_i]-4
\qR[\theta]\right)+3\qR[\theta]\ , \label{eq:CCRanomaly}
\end{align}
where $\ell(\mathbf{r})$ denotes the Dynkin index of an $\text{SU}(3)$
representation $\mathbf{r}$, normalised such that $\ell({\bf
  r}_{\text{fund}})=\frac{1}{2}$ for the fundamental representation.
Similarly, we obtain the anomaly coefficients for the remaining mixed anomalies
\begin{align}
A_{\text{SU}(2)_W-\text{SU}(2)_W-\text{U}(1)_R} &= \frac{1}{2}\left[ \qR[\Hu]+\qR[\Hd]-2\qR[\theta]+\sum_i
  \left(3\qR[\Q_i]+\qR[\Ld_i]-4\qR[\theta]\right)
  \right]+2\qR[\theta]\ ,\label{eq:WWRanomaly}\\
A_{\text{U}(1)_Y-\text{U}(1)_Y-\text{U}(1)_R}&= 2 \Bigg[\qR[\Hu]+ \qR[\Hd]-2\qR[\theta]
+\frac{1}{3}\sum_i\left(
\qR[\Q_i]+8\qR[\U_i]+2\qR[\D_i]+3\qR[\Ld_i]+6\qR[\E_i]-20\qR[\theta]\right)\Bigg]Y_L^2\ . \label{eq:YYRanomaly}
\end{align}
The three mixed anomaly coefficients of \cref{eq:CCRanomaly,eq:WWRanomaly,eq:YYRanomaly} must satisfy the Green-Schwarz anomaly
cancellation condition
\be
\label{eq:GSconditions}
\frac{A_{\text{SU}(3)_C-\text{SU}(3)_C-\text{U}(1)_R}}{k_C}
~=~\frac{A_{\text{SU}(2)_W-\text{SU}(2)_W-\text{U}(1)_R}}{k_W}
~=~\frac{A_{\text{U}(1)_Y-\text{U}(1)_Y-\text{U}(1)_R}}{k_Y}
\ ,
\ee
where $k_{...}$ denotes the Kac-Moody level of the respective gauge symmetry. Assuming gauge coupling unification in the
context of string theory~\cite{Ginsparg:1987ee}, these are related
as~\cite{Dreiner:2003yr} 
\be
k_C=k_W = \frac{3}{5}\cdot \frac{k_Y}{4 \!\!\:\cdot \!Y_L^2}  \ .\label{eq:unific}
\ee
Combining \cref{eq:GSconditions,eq:unific}, this yields two conditions
which allow us to derive new constraints on the $\UR$ charge
assignments. Using the phenomenological constraints obtained in
\cref{sec:R-FN}, we can re-express $\qR[\Q_1]$ and $\qR[\Hu]$ as
\begin{align}
\label{Q1_constraint}
\qR[\Q_1]&=\frac{1}{9} \big[-3\qR[\Ld_1]-4\left(\qR[\Hu]+\qR[\Hd]\right)
  +3x+6y-\dd-\DD+30+16\qR[\theta]\big]\ ,\\
\label{Hu_constraint}
\qR[\Hu]&=-\qR[\Hd]-z+8\qR[\theta]\ .
\end{align}

In addition to the GS conditions arising from the mixed anomalies, there
exists another anomaly coefficient, quadratic in $\UR$, which severely
constrains the allowed charge assignments.\footnote{Since we are
constructing the explicit high-energy model the comments on non-linear
anomalies in \cite{Banks:1991xj} do not apply.} As a potential anomaly
cannot be absorbed by the GS mechanism, this anomaly coefficient must 
vanish identically,
\begin{align}
\displaystyle A_{\text{U}(1)_R-\text{U}(1)_R-\text{U}(1)_Y}=&-2\Bigg[\left(\qR[\Hu]-\qR[\theta]\right)^2-\left(\qR[\Hd]-\qR[\theta]\right)^2+\sum_i\bigg(\left(\qR[\Q_i]-\qR[\theta]\right)^2-2\left(\qR[\U_i]-\qR[\theta]\right)^2+\notag \\
&+\left(\qR[\D_i]-\qR[\theta]\right)^2-\left(\qR[\Ld_i]-\qR[\theta]\right)^2+\left(\qR[\E_i]-\qR[\theta]\right)^2\bigg)\Bigg]
Y_L = 0\ . \label{eq:RRYanomaly}
\end{align}
This allows us to re-express the charge $\qR[\Hd]$ as
\begin{align}
\label{Hd_constraint}
\qR[\Hd]=\frac{1}{3(14\qR[\theta]-18-3x-2z)}&\Big[ 3\qR[\Ld_1](12-16\qR[\theta]+2x+3z)+2\dd(6-8\qR[\theta]+x+z)+2\DD(3-8 \qR[\theta]+x+z)\notag\\
&+x(14\qR[\theta]-36-6x)+z(-2z-5x-12\qR[\theta])+18-18y-156\qR[\theta]+104(\qR[\theta])^2\Big]\ .
\end{align}
In summary, anomaly considerations give rise to three new constraints as stated in 
\cref{Q1_constraint,Hu_constraint,Hd_constraint}. 
The allowed charge assignments are thus determined by the integer
parameters $x$, $y$, $z$, $\dd$ and $\DD$ as well as the charges $\qR[\Ld_1]$
and $\qR[\theta]$. 

In the following section we will further reduce the number of free parameters by
demanding the discrete~$\ZR$ symmetry to arise from the breaking of the
continuous $\text{U}(1)_R$ symmetry.
Before proceeding with this central part of our paper let us however, first discuss
the scale at which the flavon acquires its \textsc{VEV}. Thanks to the anomalous
nature of the $\UR$ symmetry, the Dine-Seiberg-Wen-Witten
mechanism~\cite{Dine:1986zy,Dine:1987bq,Atick:1987gy,Dine:1987gj} 
radiatively generates a non-vanishing Fayet-Iliopoulos term proportional to
the gravitational anomaly. The latter can in turn be related to the
$A_{\text{SU}(3)_C-\text{SU}(3)_C-\text{U}(1)_R}$ anomaly using the
\textsc{GS} mechanism. The non-vanishing Fayet-Iliopoulos term then induces a
flavon \textsc{VEV} such that~\cite{Dreiner:2003yr}
\begin{align}
\epsilon=\frac{\langle \phi \rangle}{\mgrav}&=\frac{g_C}{4\pi}\sqrt{A_{\text{SU}(3)_C-\text{SU}(3)_C-\text{U}(1)_R}}=\frac{g_C}{4\pi} \sqrt{\frac{3}{2}\left(x+z-6\qR[\theta]+6\right)}, \label{eq:epsilon-value}
\end{align}
where we have calculated the colour anomaly using the constraints on the $\UR$
charges derived above.

As previously stated, in order for the \textsc{FN} mechanism to
explain the fermion mass hierarchies we must be able to identify $\epsilon$ with
the Wolfenstein parameter, that is $\epsilon \sim \lambda_c \sim
0.22$. However one should bear in mind that the \textsc{FN} mechanism
does not fix order-one coupling constants. This implies that one can allow for
values of $\epsilon$ which vary by small amounts from $\lambda_c$. 
Mindful of this and the possible choices for $x$ and $z$, we can limit the
allowed region for $\qR[\theta]$ within a small interval. Using the
\textsc{GUT} scale value of the strong coupling, $g_C\sim 0.72$, we find conservatively
\be
- 2 ~\lesssim ~    \qR[\theta]
~\lesssim ~ 1 \ . \label{eq:interv}
\ee
As will be shown in \cref{sec:z4r}, the requirement of a residual $\ZR$
symmetry constrains $\qR[\theta]$ to odd-integer multiples
of~${1}/{4}$. Together with \cref{eq:interv}, this entails six
possible values for $\qR[\theta]$, namely $\qR[\theta]=-7/4,-5/4, -3/4,-1/4, 1/4,3/4$.
We shall, however, not include all possibilities in our subsequent analyses
since many of these solutions generate a highly \textsc{FN} suppressed
$\mu$-term. The exponent $|n|=|n_a|=|\qR[\Hu]+\qR[\Hd]|$ in \cref{eq:muter} can
be determined from the anomaly condition of \cref{Hu_constraint} as a
function of $\qR[\theta]$ and $z$. With $z=0,1$, the four  cases
$\qR[\theta]=-7/4,-5/4, -3/4,3/4$ lead to a value of $|n|\geq 5$. We will
therefore only consider the two cases 
\be
\qR[\theta]=-1/4, 1/4\,.
\ee

\section{Residual $\boldsymbol{\ZR}$ symmetry}
\label{sec:z4r}

In this section we explicitly construct the U(1)$_R$ gauge
completion of the discrete $\ZR$ model.  In order to
identify \textsc{FN} models where the $\text{U}(1)_R$ symmetry is
broken to a residual $\ZR$ symmetry we must place additional
constraints on the $\text{U}(1)_R$ charge assignments. To do so we
consider the charge of the most general operator, {\it with possible
  factors of the flavon field removed},
\begin{align}
\label{eq:Rtotalcharge}
\qR[\text{total}] &= n_{\Hu} \qR[\Hu]+n_{\Hd} \qR[\Hd]
+ \sum_i\left(n_{\Q_i} \qR[\Q_i] + n_{\U_i} \qR[\U_i] + n_{\D_i} \qR[\D_i]
+n_{\Ld_i} \qR[\Ld_i] +n_{\E_i} \qR[\E_i]+n_{\N_i}\qR[\N_i]\right).
\end{align}
Here $i=1,2,3$ label the generations of fermions and $n_S$ denotes the number
of a specific superfield $S$ occurring in a given operator. For example the
operator $\Q_i \Hd \D_j$ has $n_{\Q_i}=1$, $n_{\Hd}=1$ and $n_{\D_j}=1$. Note
that we have included three right-handed neutrinos $\N_i$ in
\cref{eq:Rtotalcharge} in order to generate neutrino masses via the
type~I seesaw mechanism~\cite{Minkowski:1977sc,ramond-seesaw,yanagida-seesaw,Mohapatra:1979ia}. 
We point out that the presence of the standard model neutral $\N_i$ does not
alter the anomaly constraints derived in \cref{sec:anomalies}, i.e.
\cref{Q1_constraint,Hu_constraint,Hd_constraint}.  Neither
are the results of Section~\ref{sec:R-FN} affected by adding right-handed
neutrinos. Beyond the relations of
\cref{eq:Qgenrelation,eq:Dgenrelation,eq:Ugenrelation,eq:Egenrelation,eq:d-rel,eq:e-rel,eq:u-rel,eq:mu-n} we get
\be
\label{eq:n-rel}
\qR[\Ld_i]+\qR[\Hu]+\qR[\N_j]-2\qR[\theta] ~\in~ \mathbb N \ ,
\qquad\quad \qR[\N_i]-\qR[\N_j] ~\in~ \mathbb Z \ , 
\ee 
from demanding the Dirac neutrino mass terms $\Ld_i \Hu \N_j$ to be
generated in the superpotential (the second relation follows
  from the first).\footnote{Generating the Dirac mass term in the
  K\"ahler potential entails a Dirac mass matrix $M_D^{ij} \sim
  \frac{v_u\, m_{3/2}}{\mgrav}
  \epsilon^{|\qR[\Ld_i]+\qR[\Hu]+\qR[\N_j]|}$. In order for the seesaw
  formula $ M_\nu = M_D M_R^{-1} M_D^T$ to apply $ \text{max}\,(M_D)
  \ll \text{min}\,(M_R)$ must hold. This entails that $M_\nu < M_D <
  \frac{v_u\, m_{3/2}}{\mgrav}$, and an extraordinarily large soft
  breaking scale $m_{3/2}$ of at least $ \SI{500}{\TeV}$, see
  also~\cite{Dreiner:2007vp}. Therefore, such a situation is highly
  unnatural, and we do not consider it any further.}  Without loss
of generality we can take $\qR[\N_3]\leq\qR[\N_2] \leq \qR[\N_1]$.

Inserting all the constraints on the $\UR$ charge assignments, we can
re-express $\qR[\text{total}]$ in terms of $x$, $y$, $z$, $\dd$,
$\DD$, $\qR[\Ld_1]$, $\qR[\theta]$, a few irrelevant integers from
\cref{eq:n-rel} and the numbers $n_{S_i}$ which specify the number of
times $S_i$ occurs in a given operator. Having specified an operator
by the sets of integers $n_{S_i}$, it is clear that not
every term is allowed by the standard model gauge symmetry. Enforcing
standard model gauge invariance translates into the following three
constraints~\cite{Dreiner:2003yr}
\begin{align}
\mathrm{SU(3)}:\ \ \ \ \ \ \ \ \  3\Cs &= \sum_i \left(n_{Q_i}- n_{\bar{D}_i} -  n_{\bar{U}_i} \right),\label{eq:gaug1}\\
\mathrm{SU(2)}: \ \ \ \ \ \ \, \  2 \Ws&= n_{H_u}+n_{H_d} + \sum_i \left(n_{Q_i} + n_{L_i} \right) ,\label{eq:gaug2}\\
\mathrm{U(1)}_Y:\ \ \ \ \ \ \ \ \ \ \ \,
 0&=3 n_{H_u}-3 n_{H_d}  + \sum_i
\left(n_{Q_i} -4 n_{\bar{U}_i}+2 n_{\bar{D}_i}  -
3 n_{L_i}+6 n_{\bar{E}_i}\right),\label{eq:gaug3}
\end{align}
where $\mathcal C,\mathcal W \in \mathbb Z$.
Together with the constraints on the charge assignments obtained in
\cref{sec:R-FN}, these gauge group constraints allow us to eliminate
$n_{\Q_i}$, $n_{\Ld_i}$, $n_{\D_i}$  and $n_{\E_i}$ from
\cref{eq:Rtotalcharge}. Then the total $\UR$ charge of 
the most general \textsc{SM} neutral operator can be written as
\begin{align}\label{eq:Rtotalwithgauge}
 \qR[\text{total}] = &\Cs \left[
   3\qR[\Q_1]+\qR[\Ld_1]+2\left(\qR[\Hd]-\qR[\Ld_1]\right)-4\qR[\theta]\right]
-2 n_{\Hu} \qR[\theta]
+n_{\Hd} \left[\qR[\Hd]-\qR[\Ld_1]\right]
\notag\\
&- \Ws \left[
\qR[\Hd]- \qR[\Ld_1]-2 \qR[\theta]\right]
 +\sum_i
 \left(n_{\U_i}+n_{\N_i}\right)\left[\qR[\Hd]-\qR[\Ld_1]+2\qR[\theta]\right]+\mathbb{Z}\ ,
\end{align}
where $\mathbb Z$ indicates an integer which can  be compensated by
appropriate powers of the flavon field. 

So far we have not yet imposed the condition that the $\text{U}(1)_R$ symmetry
gets broken down to the residual discrete~$\ZR$ symmetry of
\cref{tab:z4chargeassignments}. 
\begin{table}
\begin{center}
\begin{tabular}{c c c c c c c c c c}
\toprule & $\phantom{\Big|}\Q_i$ & $\U_i$ & $\D_i$ & $\Ld_i$ & $\E_i$ &$\N_i$ & $\Hu$ & $\Hd$& $\theta$ \\ \midrule
$\phantom{\Big|}\ZR$ &$1$ & $1$ & $1$ & $1$ & $1$ & $1$ &$0$ & $0$ & $1$  \\\bottomrule
\end{tabular}\end{center}
\caption{The charges of the \textsc{MSSM} superfields and $\theta$ under the
  discrete $\ZR$ symmetry.} 
\label{tab:z4chargeassignments}
\end{table}
To this end, in the next step we demand that {\it each} operator allowed
(forbidden) by $\ZR$  be allowed (forbidden) at the level of the
broken $\UR$ symmetry. The explicit constraint for the latter
is given below in \cref{eq:U1Rconstraint}. 
Considering $R$-symmetries, it is necessary to discuss operators arising
in both the superpotential as well as the K\"ahler potential. With the charges
given in Table~\ref{tab:z4chargeassignments}, the $\ZR$ transformation
property of any given operator can be parametrised as
\begin{align}
\label{eq:discretenumberoffields}
\mathcal N\equiv \sum_i \left(n_{Q_i}+n_{\bar{U}_i}+n_{\bar{D}_i}+n_{L_i}+n_{\bar{E}_i}+n_{\bar{N}_i}\right) &=\left\{
 \begin{array}{l l}
4 \dsi+2+\dsc \ ,& \quad\text{for superpotential operators\ ,} \\[2mm]
4\dsi+\dsc \ ,&\quad \text{for K\"ahler potential operators\ ,}
\end{array}\right.\notag\\[2mm]
&= 4\dsi+2\kosp+\dsc\ ,
\end{align}
where $\dsi$ denotes an integer, and $S=1$ for a term in the
superpotential while $S=0$ for a term in the K\"ahler potential. The integer
$\dsc$ can then be chosen to take values between $0$ and $3$ and indicates whether
a term is allowed ($\dsc=0$) or forbidden ($\dsc=1,2,3$) by the $\ZR$ symmetry.
Combining \cref{eq:discretenumberoffields} with
\cref{eq:gaug1,eq:gaug2,eq:gaug3}, it is possible to
eliminate $n_{\Hd}$ from \cref{eq:Rtotalwithgauge}. As a result we can
re-express the total charge of an operator as
\begin{align}\label{eq:Rtotalfinal}
\qR[\text{total}]=\Cs \left[3\qR[\Q_1]+\qR[\Ld_1]\right] &+\left\{ 2\left( \Ws+\sum_i \left(n_{\U_i}+n_{\N_i}\right)-n_{\Hu}-2\dsi -\kosp\right)-\dsc \right\}\left[\qR[\Hd]-\qR[\Ld_1]\right]\notag\\
&+\left(\Ws+\sum_i
\left(n_{\U_i}+n_{\N_i}\right)-n_{\Hu}-2\Cs\right)2\qR[\theta]+\mathbb{Z}\ .
\end{align}
\cref{eq:Rtotalfinal} is our master equation which we scrutinise
in the rest of this section. 
The integers $\Cs$, $\Ws$, $n_{\U_i}$, $n_{\N_i}$, $n_{\Hu}$,
$\dsi$ as well as $\dsc$ and $S$ parametrise a particular 
operator of the superpotential ($S=1$) or K\"ahler potential ($S=0$), which
for $\dsc=0$ ( $\dsc=1,2,3$) is allowed (forbidden) under $\ZR$. 
Choosing special sets of these parameters, we discuss classes of
operators and compare their compatibility with being allowed (for $\dsc=0$)
or forbidden (for $\dsc=1,2,3$) at the level of the broken $\UR$ symmetry,
where the condition for being allowed is
\begin{align}\label{eq:U1Rconstraint}
\qR[\text{total}]&= \left\{
\begin{array}{l l}
\mathbb{N}+2\qR[\theta] \ ,& \quad\text{for superpotential operators\ ,} \\[2mm]
\mathbb{Z} \ ,&\quad \text{for K\"ahler potential operators \ .}
\end{array}\right.
\end{align}
This comparison provides constraints on the $\UR$
charges $\left[3\qR[\Q_1]+\qR[\Ld_1]\right]$, $\left[\qR[\Hd]-\qR
  [\Ld_1]\right]$ and $\qR[\theta]$ which will allow us to formulate
simple conditions for $\ZR$ to arise as a residual symmetry of the
Froggatt-Nielsen $\UR$.  In the following we focus on the class of
operators which satisfy 
\be
\Ws+\sum_i \left(n_{\U_i}+n_{\N_i}\right)-
n_{\Hu}=\kosp+2\dsi\,.
\ee
 Their total $\UR$ charge is given by 
\be
\label{eq:Rtotaltilde}
\wt R_{\text{total}} [\Cs,\dsi,\kosp,\dsc] ~=~ \Cs
\left[3\qR[\Q_1]+\qR[\Ld_1]\right] -\dsc
\left[\qR[\Hd]-\qR[\Ld_1]\right] +4 \left(\dsi - \Cs \right)
\qR[\theta] + 2 \kosp\qR[\theta] +\mathbb{Z}\ .  
\ee 
We first study cases with $\Cs=0$ to derive constraints on
$\qR[\theta]$ and $\left[\qR[\Hd]-\qR[\Ld_1]\right]$. Equipped with
these results, we then consider $\Cs=1$ in order to constrain
$\left[3\qR[\Q_1]+\qR[\Ld_1]\right]$. 
\begin{itemize}
\item[$(i)$] \underline{$\Cs=0$:} A $\ZR$ invariant operator has $\dsc=0$ and
  a total $\UR$ charge of  
\be
\wt R_{\text{total}} [\Cs=0,\dsi,\kosp,\dsc=0]
~=~
4 \dsi  \qR[\theta] + 2 \kosp\qR[\theta]
+\mathbb{Z}\ .
\ee
Since we want complete agreement between the $\ZR$ symmetry of
Table~\ref{tab:z4chargeassignments} and the remnant symmetry after $\UR$
breaking, this equation must hold for all $\dsi$. Comparing the above for
arbitrary $\dsi$ with the condition for a $\UR$ allowed operator, see
\cref{eq:U1Rconstraint}, it is clear that
\begin{align}
\qR[\theta]=\frac{\mathbb{Z}}{4} \ ,\label{eq:theta-pre}
\end{align}
must be satisfied. This requirement removes the third term
in \cref{eq:Rtotaltilde}, that is the one proportional to $(\dsi-\Cs)$. With
this simplification, we consider $\ZR$ violating 
operators with $\dsc=1$ and a total $\UR$ charge of
\be
\wt R_{\text{total}} [\Cs=0,\dsi,\kosp,\dsc=1]
~=~
- \left[\qR[\Hd]-\qR[\Ld_1]\right]
+ 2 \kosp\qR[\theta]
+\mathbb{Z}\ .
\ee
Demanding that such a term be forbidden at the $\UR$ level,
tells us that $\left[\qR[\Hd]-\qR[\Ld_1]\right] \notin
\mathbb{Z}$. In order to extract further or more specific information
on the charge assignments, it is necessary to consider various powers
of the $\ZR$ violating operators with $\dsc=1$. From
\cref{eq:discretenumberoffields}, one finds that the square of a
(forbidden) superpotential term with $\dsc=1$ will be allowed in the
superpotential due to its resulting $\ZR$ charge of~2. In contrast,
only the fourth power of a (forbidden) K\"ahler potential term with
$\dsc=1$ will be allowed in the K\"ahler potential, while the square
and the cube will still be forbidden by the $\ZR$
symmetry. Translating these observations to constraints on the $\UR$
charges we get
$$
\begin{array}{cccl}
 &2 \left[\qR[\Hd]-\qR[\Ld_1]\right]= 2\qR[\theta] + \mathbb Z \ ,  &&
 ~~~~\text{from superpotential operators\ ,} \\[2mm]
 2 \left[\qR[\Hd]-\qR[\Ld_1]\right]\neq  \mathbb Z \ , &
3 \left[\qR[\Hd]-\qR[\Ld_1]\right]\neq  \mathbb Z \ ,  &
 4 \left[\qR[\Hd]-\qR[\Ld_1]\right]=  \mathbb Z \ , & ~~~~\text{from K\"ahler
   potential operators\ .}
\end{array}
$$
Consistency of these conditions requires $2\qR[\theta]\notin
\mathbb{Z}$. Together with \cref{eq:theta-pre}, the most general choice
for $\qR[\theta]$ consistent with a residual $\ZR$ symmetry is an odd-integer
multiple of $\frac{1}{4}$, and can be written as
\be
\qR[\theta]=\frac{2\,\mathbb{Z}+1}{4} \ .\label{eq:thetaA}
\ee
This in turn entails that $\left[\qR[\Hd]-\qR[\Ld_1]\right]$ must also be an odd-integer
multiple of $\frac{1}{4}$, hence 
\begin{align}\label{eq:KpHdLdfinalsolution}
\left[\qR[\Hd]-\qR[\Ld_1]\right] &= \frac{2\,\mathbb{Z}+1}{4}\ .
\end{align}
\item[$(ii)$] \underline{$\Cs=1$:} 
We proceed with determining the allowed values of
$3\qR[\Q_1]+\qR[\Ld_1]$. We reconsider \cref{eq:Rtotaltilde}, this time
setting $\Cs=1$. The $\UR$ charge of a  $\ZR$ invariant operator is given by
$$
\wt R_{\text{total}} [\Cs=1,\dsi,\kosp,\dsc=0]
~=~
\left[3\qR[\Q_1]+\qR[\Ld_1]\right] 
 + 2 \kosp\qR[\theta]
+\mathbb{Z}\ .
$$
Comparison with \cref{eq:U1Rconstraint} immediately reveals that 
\begin{align}
\left[ 3\qR[\Q_1]+\qR[\Ld_1]\right]=\mathbb{Z}\ .  \label{eq:3q-l}
\end{align}
\end{itemize}
We have  constrained all three parameters in Eq.~(\ref{eq:Rtotaltilde}): $\qR[\theta]$,
$\left[\qR[\Hd]-\qR[\Ld_1]\right]$ and $\left[ 3\qR[\Q_1]+\qR[\Ld_1]\right]$
via \cref{eq:thetaA,eq:KpHdLdfinalsolution,eq:3q-l}
using particular classes of operators. It is straightforward to show that the
re-insertion of these values into \cref{eq:Rtotalfinal} ensures that
\cref{eq:U1Rconstraint} is fulfilled for all $\ZR$ allowed ($\dsc=0$)
operators while it is not satisfied for all $\ZR$ forbidden ($\dsc=1,2,3$)
operators, regardless of the values of the free integer parameters 
 $\Cs$, $\Ws$, $n_{\U_i}$, $n_{\N_i}$, $n_{\Hu}$ and 
$\dsi$. 

Using notation similar to Refs.~\cite{Dreiner:2003yr,Dreiner:2006xw,Dreiner:2007vp}
we re-express the constrained parameters as, 
\begin{align}
\left[\qR[\Hd]-\qR[\Ld_1]\right]&=\frac{2\mathbb{Z}+1}{4} 
\qquad \longrightarrow \qquad
 \Dh  \equiv 2\left[\qR[\Ld_1]-\qR[\Hd]\right]-\frac{1}{2}\ ,\\
\left[3\qR[\Q_1]+\qR[\Ld_1]\right]&=\mathbb{Z} 
\hspace{8.55mm} 
\qquad \longrightarrow \qquad
\hspace{2.33mm}
3\squi  \equiv \dd+\DD-z+4\qR[\theta]\ ,\label{eq:3Q1Ld1rep}
\end{align}
where both $\Dh$ and $\squi$ are independent integers, and
\cref{eq:3Q1Ld1rep} is obtained through the use of
\cref{Q1_constraint,Hu_constraint}, namely, 
\begin{align}\label{eq:squiderivation}
3\qR[\Q_1]+\qR[\Ld_1]&=10-\frac{1}{3}\left(\dd+\DD\right)+x+2y+\frac{4}{3}z-\frac{4}{3}\left(4\qR[\theta]\right)~\stackrel{!}{=}~\mathbb{Z}\ .
\end{align}
With $x$, $y$, $z$ and $4\qR[\theta]$ being integers, 
the combination $\dd+\DD-z+4\qR[\theta]$ must be an integer multiple of 3.

We collect all the information on the constrained charge assignments
derived in this and the previous sections in \cref{tab:table1}. The
$\UR$ charges of the standard model charged fields depend on
the parameters $x$, $y$, $z$, $\DD$, $\Dh$, $\zeta$ and $\qR[\theta]$,
where the latter has to be an odd-integer multiple of
$\frac{1}{4}$. The charges of the right-handed neutrinos must be
consistent with \cref{eq:n-rel}, but remain otherwise unconstrained at
this point. We discuss this in more detail in the next section.
\begin{table}
\begin{center}
\begin{tabular}{c}
\toprule 
\parbox{10cm}{
\begin{align*} 
R_{H_d} &=\frac{1}{20 (6-6 \qR[\theta]+x+z)}
\Big[
(46-4 \Dh+8x+4z)x
+24 y-(19+6 \Dh) z\\&
-8 (6+x+z) \squi 
-4 (9-2  \DD+6 \Dh)
+(288+32 \Dh-8 x+48z+64\squi ) \qR[\theta] 
-224 (\qR[\theta])^2
\Big] \\
\qR[\Hu]&=-\qR[\Hd]-z+8\qR[\theta]\\
\qR[\Q_1]&=\frac{1}{3}\left[\frac{39}{4}-\qR[\Hd]+x+2y+z-\squi -\frac{\Dh}{2}-4\qR[\theta]\right]\\
\qR[\Q_2]&=\qR[Q_1]-1-y\\
\qR[\Q_3]&=\qR[\Q_1]-3-y\\
\qR[\U_1]&=-\qR[\Q_1]-\qR[\Hu]+8+2\qR[\theta]\\
\qR[\U_2]&=\qR[\U_1]-3+y\\
\qR[\U_3]&=\qR[\U_1]-5+y\\
\qR[\D_1]&=-\qR[\Q_1]-\qR[\Hd]+4+x+2\qR[\theta]\\
\qR[\D_2]&=\qR[\D_1]-1+y\\
\qR[\D_3]&=\qR[\D_1]-1+y\\
\qR[\Ld_1]&= \qR[\Hd]+\frac{1}{4}\left(2\Dh+1\right)\\
\qR[\Ld_2]&=\qR[\Ld_1]-\DD+z+3\squi-4\qR[\theta]\\
\qR[\Ld_3]&=\qR[\Ld_1]+\DD\\
\qR[\E_1]&=-\qR[\Ld_1]-\qR[\Hd]+4+x+z+2\qR[\theta]\\
\qR[\E_2]&=\qR[\E_1]+\DD-2-2z-3\squi+4\qR[\theta]\\
\qR[\E_3]&=\qR[\E_1]-\DD-4-z
\end{align*}}\\
\bottomrule
\end{tabular}\end{center}
\caption{The $\UR$ charges of the standard model charged matter fields. These are
  constrained by the low-energy phenomenology of quark masses, charged lepton
  masses and quark mixing, the \textsc{GS} anomaly conditions as well as the
  requirement of producing the residual $\ZR$ symmetry. The free parameters
  are $x=0,1,2,3$, $y=-1,0,1$, $z=0,1$, the integers $\DD$, $\Dh$, $\zeta$ as
  well as $\qR[\theta]$ which must be an odd-integer multiple of
  $\frac{1}{4}$.}
\label{tab:table1}
\end{table}

\section{Neutrino constraints}
\label{sec:neutrino}

In this section, we discuss the possibilities of
introducing neutrino masses to the model which  are
compatible with the residual $\ZR$ symmetry. As $\ZR$ includes
$R$-parity, the only way to generate neutrino masses in the absence of
right-handed neutrinos is through the effective Weinberg operator
$\Ld_i \Hu \Ld_j \Hu$~\cite{Weinberg:1979sa}.  Being generated
gravitationally, this term gives rise to a neutrino mass matrix of the
form
\be 
M_{\nu}^{ij}
~\sim~ \frac{v_u^2}{\mgrav}\cdot
\left(\frac{m_{3/2}}{\mgrav}\right)^{1-S} \cdot ~
\epsilon^{|\qR[\Ld_i]+\qR[\Ld_j]+2\qR[\Hu]-2S\qR[\theta]|}\
,\label{eq:weinb} 
\ee 
where $S=1$ if it originates in the superpotential, and $S=0$ if it
arises from the K\"ahler potential. As the \textsc{FN} expansion
parameter $\epsilon$ can only lead to further suppression, the
absolute neutrino mass scale is bounded from above by the ratio
between the square of the Higgs \textsc{VEV} and the gravitational
scale. With $v_u\sim m_t \sim \SI{175}{\GeV}$ and $\mgrav \sim
\SI{2.4e18}{\GeV}$, we obtain the maximum mass scale of the neutrinos
to be of order $\SI{e-5}{\eV}$ if the term originates in the
superpotential and even smaller if the Weinberg operator is generated
in the K\"ahler potential. This result is in conflict with atmospheric
neutrino oscillations observations which require the absolute neutrino
mass scale to be bigger than $\SI{e-2}{\eV}$. Hence, we do not consider this
option any further.

\subsection{Seesaw mechanism}
To account for realistic neutrino masses we introduce three right-handed
neutrinos $\N_i$ and impose the type~I seesaw mechanism~\cite{Minkowski:1977sc,ramond-seesaw,yanagida-seesaw,Mohapatra:1979ia}. 
This mechanism relies on a Majorana mass term
\be
M_{R}^{ij} \N_i \N_j\ ,\label{eq:rhmass}
\ee
for the right-handed neutrinos as well as a Dirac mass term
\be
Y_{D}^{ij} \Ld_i \Hu \N_j\ ,\label{eq:diracmass}
\ee
which couples left- and right-handed neutrinos. 
Each of these terms can originate in either the superpotential
(requiring $\qR[\text{total}]=2\qR[\theta]+\mathbb N$) or the K\"ahler
potential (requiring $\qR[\text{total}]=\mathbb Z$).  However, we argued in
\cref{sec:z4r} that the Dirac term should arise in the superpotential, and
this requirement is in fact built into the charge assignments listed in
\cref{tab:table1}. As a consequence, we now show, the Majorana mass term
cannot originate from the  K\"ahler potential. Consider the
  square of the Dirac operator in \cref{eq:diracmass}. Since $\qR[\theta]$
  is an (odd-)integer multiple of~$\frac{1}{4}$, see \cref{eq:thetaA},
  the squared Dirac term must have integer total $\UR$ charge, i.e.
  $\qR[\Ld_i]+\qR[\Ld_{j}]+2\qR[\Hu] +\qR[\N_{i'}] +\qR[\N_{j'}] \in
  \mathbb Z$.  However, using the $\UR$ charges of \cref{tab:table1}
  we have $ \qR[\Ld_i]+\qR[\Ld_j]+2\qR[\Hu] ~=~ \frac{1}{2} + \mathbb
  Z$, so that
\be
\qR[\N_{i}] +\qR[\N_{j}] ~=~ \frac{1}{2} + \mathbb Z \ .\label{eq:help}
\ee
Hence, the Majorana mass term of \cref{eq:rhmass} can only arise
from the superpotential, cf. Eq.~(\ref{eq:U1Rconstraint}).

Thus both the Majorana and the Dirac mass terms are generated in the
superpotential. The corresponding mass matrices are then of the form
\be
M_R^{ij} ~\sim~ \mgrav ~
 \epsilon^{\qR[\N_i]+\qR[\N_j]-2\qR[\theta]} \ , \qquad \qquad
M_D^{ij} ~\sim~ v_u~
 \epsilon^{\qR[\Ld_i]+\qR[\Hu]+\qR[\N_j]-2\qR[\theta]} \ , \label{eq:massmatrices}
\ee
where the exponents must be non-negative integers for any choice of
$i,j=1,2,3$.  Due to the orderings 
$\qR[\N_3]\leq \qR[\N_2] \leq \qR[\N_1]$ and
$\qR[\Ld_3]\leq \qR[\Ld_2] \leq \qR[\Ld_1]$, it is sufficient to impose
\be
2 \qR[\N_3]-2\qR[\theta] ~\in~ \mathbb{N}\ , \quad\qquad
 \qR[\Ld_3]+\qR[\Hu]+\qR[\N_3] -2\qR[\theta] ~\in~ \mathbb{N} \ .\label{eq:Nconstr1}
\ee
Assuming only moderate $\epsilon$~suppressions, 
the Majorana and Dirac masses take values around their natural scale, and the
seesaw formula is applicable as  long as $
\text{max}\,(M_D) \ll  \text{min}\,(M_R)$, i.e. 
\be
\frac{v_u}{\mgrav} ~<~ \epsilon^{2\qR[\N_1]-\qR[\N_3]-\qR[\Ld_3]-\qR[\Hu]} \ .\label{eq:Nconstr2}
\ee
Inserting the matrices of \cref{eq:massmatrices} into the seesaw formula
$
M_\nu = M_D M_R^{-1} M_D^T
$
yields an effective light neutrino mass matrix of the form
\be
M_{\nu}^{ij} ~\sim~  
\frac{v_u^2}{\mgrav}
 \epsilon^{\qR[\Ld_i]+\qR[\Ld_j]+2\qR[\Hu]-2\qR[\theta]}\ ,\label{eq:FNneutrinoma}
\ee
where, in contrast to \cref{eq:weinb}, the exponent  $\qR[\Ld_i]
+\qR[\Ld_j]+2\qR[\Hu]-2\qR[\theta]$ can now take negative integer
values. Hence, it is possible to end up with a phenomenologically acceptable
neutrino mass scale above $\SI{e-2}{\eV}$. In the next subsection we study
\cref{eq:FNneutrinoma} with the aim of extracting further constraints on
the possible $\UR$ charge assignments.

\subsection{Neutrino masses and mixings}

The neutrino mass matrix of \cref{eq:FNneutrinoma} can  be diagonalised
by a unitary matrix $\wt V_{\nu_L}^{ij} \sim \epsilon^{|\qR[\Ld_i]- \qR[\Ld_j]|}$
which has the same $\epsilon$~structure as the left-chiral unitary
transformation $V_{e_L}^{ij}$ required to diagonalise the charged lepton mass
matrix. 
The singular values $\wt m_i$ resulting from diagonalising
\cref{eq:FNneutrinoma} by $\wt V_{\nu_L}^{ij}$ are related as
\be
\wt m_1:\wt m_2:\wt m_3 ~\sim~1:\epsilon^{2\dd}:\epsilon^{2\DD}\ 
. \label{eq:singularratios}
\ee
As the ordering of the neutrino masses is still unknown, the
Pontecorvo-Maki-Nakagawa-Sakata (PMNS) matrix 
can be written as
\be
U_{\textsc{PMNS}} = V_{e_L}\wt V^\dagger_{\nu_L}  T \ ,\label{eq:pmnsdef}
\ee
where $T$ indicates a matrix which permutes the singular values $\wt m_i$
appropriately to make it consistent with either a normal or an inverted mass
ordering. In accordance with
Refs.~\cite{Tortola:2012te,Fogli:2012ua,GonzalezGarcia:2012sz}, we can now  
parametrise the experimentally observed PMNS matrix in the following manner, 
\begin{align}\label{eq:MSN}
U_{\textsc{PMNS}} \sim \begin{pmatrix}
\epsilon^{0,1} & \epsilon^{0,1} & \epsilon^{0,1} \\
\epsilon^{0,1} & \epsilon^{0,1} & \epsilon^{0,1} \\
\epsilon^{0,1} & \epsilon^{0,1} & \epsilon^{0,1} 
\end{pmatrix},
\end{align}
where the $0,1$ indicates the allowed exponents.\footnote{We remark that 
more accurate or special patterns of lepton mixing can only be explained by
means of non-Abelian family symmetries with triplet representations~\cite{Luhn:2007yr,Ludl:2009ft,Altarelli:2010gt,Ishimori:2010au,Grimus:2011fk,King:2013eh}.} Due to
the democratic pattern of \cref{eq:MSN},  $U_{\textsc{PMNS}} T^{-1}$ and
$U_{\textsc{PMNS}}$ have identical $\epsilon$ structure, and \cref{eq:pmnsdef}
entails 
$
\epsilon^{0,1} \sim \epsilon^{|\qR[\Ld_i]-\qR[\Ld_j]|}
$.
Thus, the PMNS mixing constrains the $\UR$ charges $\qR[\Ld_i]$ such that only
three different combinations of the parameters $(\dd,\DD)$ consistent with 
the charge ordering $\DD\leq \dd\leq 0$ are allowed
\be
(\dd,\DD) ~=~(0,0)~~\text{or}~~(0,-1)~~\text{or}~~(-1,-1)\ .\label{eq:deltaconds}
\ee
Disregarding the ordering of the masses in \cref{eq:singularratios}, which
can be corrected by the permutation matrix $T$, the three possible choices for
$(\dd,\DD)$ are compatible with different neutrino mass spectra as follows:
\begin{itemize}
\item $(0,0)$ allows for normal hierarchy, inverted hierarchy and degenerate neutrinos,
\item $(0,-1)$ allows for normal hierarchy,
\item $(-1,-1)$ allows for normal hierarchy and inverted hierarchy.
\end{itemize}

Turning to the absolute neutrino mass scale, it is given by
$M_\nu^{33}$ of \cref{eq:FNneutrinoma}. 
With the charges listed in \cref{tab:table1}, we find
\begin{align}
m_{\text{abs}}^\nu \sim \frac{v_u^2}{\mgrav}\epsilon^{\Dh-2z+\frac{1}{2} +2\DD+14\qR[\theta]}.\label{eq:absol}
\end{align}
Solving for the exponent yields
 \begin{align}
\mathcal Z ~\equiv~\Dh-2z+\frac{1}{2} +2\DD+14\qR[\theta]
&~\sim ~ \frac{1}{\ln{\epsilon}} \ln \left(\frac{m_{\text{abs}}^\nu \mgrav}{v_u^2}\right).\label{eq:zcalcu}
 \end{align}
Thanks to $\qR[\theta]$  being an odd-integer multiple of $\frac{1}{4}$, the
parameter $\mathcal Z$ is an integer. Its value is further constrained by
$\epsilon$ as well as the absolute neutrino mass scale
$m_{\text{abs}}^\nu$. With three possible neutrino mass
spectra~\cite{GonzalezGarcia:2007ib}  
\begin{subequations}
\begin{align}
m_1 < m_2 \ll m_3 \sim \SI{0.05}{\eV}\ ,& \qquad \text{(normal hierarchy)\ ,}\\[2mm]
m_3 \ll m_1 < m_2 \sim  \SI{0.05}{\eV}\ ,& \qquad \text{(inverted hierarchy)\ ,}\\[2mm]
\SI{0.2}{\eV} < m_1 \sim m_2 \sim m_3 < \SI{2.2}{\eV}\ ,& \qquad
\text{(degenerate)\ ,}
\end{align}
\end{subequations}
it is necessary to consider only two different values of $m_{\text{abs}}^\nu$. To account for the variation of the unknown 
order-one coupling coefficients, we allow $m_{\text{abs}}^\nu$ to vary over a small interval. This interval is given by the multiplication and division of $\sqrt{10}$ for the upper and lower bounds of $m_{\text{abs}}^\nu$ respectively, which with small amounts of rounding yields
\begin{align}
\SI{0.015}{\eV} ~ \lesssim ~m_{\text{abs}}^\nu ~ \lesssim ~ \SI{0.16}{\eV}\ ,& \qquad \text{(normal and inverted hierarchy)\ ,}\label{eq:hiera-abs}\\[2mm]
\SI{0.06}{\eV} ~ \lesssim ~m_{\text{abs}}^\nu ~ \lesssim ~ \SI{7.00}{\eV}\ ,& \qquad
\text{(degenerate)\ .}\label{eq:degene-abs}
\end{align}
Concerning the \textsc{FN} expansion parameter, we consider only values within
the interval $0.16 \lesssim \epsilon \lesssim 0.28$. This continuous interval
is broken up into only a discrete set since \cref{eq:epsilon-value}
depends solely on a particular linear combination of the parameters
$x=0,1,2,3$, $z=0,1$ and $\qR[\theta]=\pm\frac{1}{4}$. With these constraints
on $m_{\text{abs}}^\nu$ and $\epsilon$, it is straightforward to determine the
allowed values of $\mathcal Z$ from \cref{eq:zcalcu}. The results are listed in
\cref{tab:mathcalZ-eps}. Notice that larger values of $\epsilon$ cannot
be achieved once we restrict ourselves to
$\qR[\theta]=\pm\frac{1}{4}$, see the end of \cref{sec:anomalies}. 
\begin{table}
\begin{center}
\begin{tabular}{cccc}
\toprule
$~~\epsilon~~$ & $~~x+z-6\qR[\theta]~~$ & 
~~~~$\mathcal Z_{\text{hier}}$~~~~
& ~~~~$\mathcal Z_{\text{deg}}$~~~~
\\ \midrule
$0.165$  & $-0.5$ & $-5,-4$ & $-7,-6,-5$  \\
$0.179$  & $0.5$ & $-5$ & $-7,-6,-5$  \\
$0.192$  & $1.5$ & $-5$ & $-8,-7,-6$  \\
$0.205$  & $2.5$ & $-5$ & $-8,-7,-6$  \\
$0.216$  & $3.5$ & $-6,-5$ & $-8,-7,-6$  \\
$0.227$  & $4.5$ & $-6,-5$ & $-9,-8,-7,-6$  \\
$0.238$  & $5.5$ & $-6,-5$ & $-9,-8,-7,-6$  \\ \bottomrule
\end{tabular}\end{center}
\caption{The possible values for the integer $\mathcal Z$ of
  \cref{eq:zcalcu} depending on the
  \textsc{FN} expansion parameter $\epsilon$ and the neutrino mass
  spectrum, see \cref{eq:hiera-abs,eq:degene-abs}. The third column is valid for a normal and an inverted neutrino mass
  hierarchy, while the fourth column holds for the degenerate case.}
\label{tab:mathcalZ-eps}
\end{table}

\subsection{Collection of all constraints and resulting models}
The charges of \cref{tab:table1} depend on the parameters $x$, $y$, $z$,
$\DD$, $\Dh$, $\zeta$ and $\qR[\theta]$. As argued at the end of
\cref{sec:anomalies}, we constrain ourselves to cases where
\begin{align}
\qR[\theta] &= \pm \frac{1}{4} \ .\label{eq:final1}
\end{align}
All remaining six parameters are integers which are restricted to the values
\begin{subequations}
\begin{align}
x&=0,1,2,3 \ ,\label{eq:final2}\\
y&=-1,0,1 \ ,\label{eq:final3}\\
z&= 0,1 \ ,\label{eq:final4} \\
\DD &= 0,-1 \ , \label{eq:final5}\\
\Dh&=\mathcal Z +2z- \frac{1}{2} -2\DD-14\qR[\theta] \ ,\label{eq:final6}\\
\zeta &= \frac{1}{3} \left(\dd + \DD -z+4\qR[\theta]  \right) \ ,\label{eq:discr-con}
\end{align}
\end{subequations}
where $\mathcal Z$ and $\dd$ must comply with \cref{tab:mathcalZ-eps} and
\cref{eq:deltaconds}, respectively. In particular, parameter sets for
which the right-hand side of \cref{eq:discr-con} does not yield an
integer are excluded. Moreover, as discussed below \cref{eq:deltaconds},
only certain mass spectra are allowed for a given choice of
$(\dd,\DD)$. Taking into account all these constraints, we end up with a
limited number of viable models which are defined by the sets of parameters
listed in \cref{tab:collect-final}. In addition to the free parameters defined above
the table shows the allowed neutrino mass spectra as well as the ranges for
$m_{\text{abs}}^\nu$ and $\epsilon$. Ignoring the right-handed neutrinos, we
thus have identified 3~$\times$~34~different phenomenologically acceptable
\textsc{FN} $\UR$ models which give rise to the residual $\ZR$ symmetry of
\cref{tab:z4chargeassignments}. The explicit $\UR$ charge assignments can
be found in \cref{app:tables}. 

We remark that the origin of the $\mu$-term is model dependent. As all the
parameters are fixed for a given model, Table~\ref{tab:table1} allows us to
calculate $n=n_a=\qR[\Hu]+\qR[\Hd]=-z+8 \qR[\theta]$, see Eq.~\eqref{eq:mu-n}.
Cases 1, 2, 3 and 4 of Table~\ref{tab:collect-final} (including all their
subcases) yield a negative value of $n$, so that the $\mu$-term necessarily
has to arise via the Giudice-Masiero mechanism. All the other cases of
Table~\ref{tab:collect-final} (including all their subcases) have $n>0$, so that both a
Kim-Nilles as well as a Giudice-Masiero origin are possible.
\begin{table}
\begin{center}
\begin{tabular}{rcccccccccccc}
\toprule
\# & $~R_\theta~$ & $x$ & $y$ & $~~z~~$  & $~\dd~$ & $~\DD~$ & $~\zeta~$ &
 $~\mathcal Z~$ & $~\Dh~$
  &  ~spectrum~ & ~~$m_{\text{abs}}^\nu [\text{eV}]$~~ & ~~$\epsilon$~~  \\ \midrule
1 & $-1/4$  & $~0,1,2,3~$ & $-1,0,1$ & $1$ & $0$ & $-1$ & $-1$  &
$-5$ &       $2$  & norm.  & $0.036\rightarrow 0.017$ & $0.205 \rightarrow 0.238$  \\
2 & & $1,2,3$ & &&& & & $-6$ & $1$ &  norm. & ~~$0.125\rightarrow 0.070$~~ &
~~$0.216 \rightarrow 0.238$~~ \\ \midrule
3 &$-1/4$  & $0,1,2,3$ & $-1,0,1$ & $0$ & $-1$ & $-1$ & $-1$   &
$-5$ &       $0$ & ~~norm.~\& inv.~~ & $0.049\rightarrow 0.021$ & $0.192\rightarrow 0.227$  
\\
4 && $2,3$ & &&& & & $-6$ & $-1$ & ~~norm.~\& inv.~~ & $0.125\rightarrow 0.092$ & $0.216\rightarrow 0.227$   \\ \midrule
5 & $~\phantom-1/4~$  & $0$ & $-1,0,1$ & $1$ & $0$ & $0$ & $0$  &  $-4$ &
$-6$ & ~~norm.~\& inv.~~ & $0.017$ & $0.165$
\\
6 && $0,1,2,3$ & &&& & &  $-5$  & $-7$ & ~~norm.~\& inv.~~ & $0.106\rightarrow
0.036$ & $0.165\rightarrow 0.205$ \\
7 && $0,1$ & &&& & &  $-5$  & $-7$ & ~~deg.~~ & $0.106\rightarrow
0.070$ & $0.165\rightarrow 0.179$ \\
8 && $0,1,2,3$ & &&& & & $-6$ & $-8$ & ~~deg.~~  & $0.642\rightarrow
0.174$ & $0.165\rightarrow 0.205$ \\
9 && $0,1,2,3$ & &&& & & $-7$  & $-9$ & ~~deg.~~ & $3.903 \rightarrow
0.851$ & $0.165\rightarrow 0.205$  \\
10 & & $2,3$ & &&& & & $-8$  & $-10$ & ~~deg.~~  & $6.859 \rightarrow
4.158$ & $0.192\rightarrow 0.205$ \\ \midrule
11 & $\phantom-1/4$  & $1$ & $-1,0,1$ & $0$ & $0$ & $-1$ & $0$  &  $-4$ &       $-6$ & norm.   & $0.017$ & $0.165$
\\
12 & & $1,2,3$ & &&& & &  $-5$ & $-7$ & norm. & $0.106 \rightarrow
0.049$ &  $0.165 \rightarrow 0.192$  \\ \bottomrule
\end{tabular}\end{center}
\caption{The sets of parameters that are compatible with the conditions of
  \cref{eq:final1,eq:final2,eq:final3,eq:final4,eq:final5,eq:final6,eq:discr-con,tab:mathcalZ-eps,eq:deltaconds} together with the corresponding mass spectra
allowed by the given choice of $(\dd,\DD)$. The last two columns show the
possible ranges for $m_{\text{abs}}^\nu$ and $\epsilon$. These two values are calculated from
\cref{eq:absol,eq:epsilon-value} respectively, using the
freedom in the parameter~$x$ but disregarding potential contributions from order-one
factors.  }
\label{tab:collect-final}
\end{table}

\subsection{Right-handed neutrino charges}
Before concluding this section, we wish to comment on the charges of
the right-handed neutrinos~$\N_i$.  As \cref{eq:FNneutrinoma} does not
depend on $\qR[\N_i]$, there exists no constraint from low-energy
neutrino physics. However, as discussed above, consistency requires
the charges to satisfy the conditions of
\cref{eq:Nconstr1,eq:Nconstr2}.\footnote{The condition of
  \cref{eq:help} automatically holds if one of the two constraints in
  \cref{eq:Nconstr1} is fulfilled.}  A further constraint can be
imposed if one demands successful baryogenesis via
leptogenesis~\cite{Fukugita:1986hr}. In the case of ``vanilla'' 
leptogenesis, the mass of the lightest right-handed neutrino has to be
larger than $M_{\text{lept}} \sim
\SI{e9}{\GeV}$~\cite{Davidson:2002qv,Buchmuller:2002rq,Buchmuller:2004nz},
which translates into 
\be M_{\text{lept}} ~\lesssim ~ \mgrav ~
\epsilon^{2\qR[\N_1]-2\qR[\theta]} \ .  
\ee 
Since the mass scale of the Dirac neutrino term is necessarily smaller
or equal to the Higgs VEV $v_u\sim \SI{175}{\GeV}$, the constraint of
\cref{eq:Nconstr2} is always satisfied. We are then left with the conditions of
\cref{eq:Nconstr1} 
\bea
2 \qR[\N_3]-2\qR[\theta] &~\in~& \mathbb{N}\ , \\
\qR[\N_3] +\frac{\mathcal Z}{2}-\qR[\theta] &~\in~& \mathbb{N} \ ,
\eea
yielding a lower bound on $\qR[\N_3]$, as well as the leptogenesis constraint
\be
2\qR[\N_1]-2\qR[\theta] ~ \lesssim ~\frac{1}{\ln \epsilon} ~ \ln\left(
\frac{M_{\text{lept}}}{\mgrav} \right) \ ,
\ee
and hence an upper bound on $\qR[\N_1]$. With the ordering $\qR[\N_3]\leq
\qR[\N_2] \leq \qR[\N_1]$ it is now straightforward to count the number of
all distinct allowed right-handed neutrinos charge assignments for each model
in \cref{tab:collect-final}.

\section{Conclusion}
\label{sec:conclusion}

In this paper we have successfully constructed a viable set of flavour
models  with a gauged $\text{U(1)}_R$ family symmetry. We have determined
these models such that they act as a $\UR$ gauge completion of the $\ZR$ symmetry. 
Through the use of the Green-Schwarz mechanism for anomaly cancellation
and the insistence of the residual $\ZR$ symmetry after $\text{U(1)}_R$
breaking we are able to constrain the free parameters of the theory. 
Furthermore, if one insists on phenomenologically acceptable quark and
lepton masses and mixings at the \textsc{GUT} scale, as well as the inclusion of
neutrino masses using the type~I seesaw mechanism, we obtain unique
sets of charge assignments under the original $\text{U(1)}_R$
symmetry. Although the right-handed neutrino charges remain undetermined, 
they are severely constrained by the required assumptions for a successful
type~I seesaw mechanism, as well as by the minimum right-handed neutrino mass
required for successful leptogenesis.

Additionally we have discussed the origin of the $\mu$-term in these
models. We insist that it remains forbidden under the $\text{U(1)}_R$
symmetry and is dynamically generated during the breaking of $\ZR$ to
$\mathbb{Z}_2^R$ through either the Kim-Nilles or the Giudice-Masiero
mechanism. While both mechanism have different origins the
results in the context of the models presented are indistinguishable.

The result of this work is a set of 3~$\times$~34 models presented in
\cref{app:tables}. Adopting a $\UR$ charge normalisation where the flavon
field has charge $\qR[\phi]=-1$, most models suffer from highly fractional
charges. The most aesthetically pleasing ones are
\begin{itemize}
\item case 3a of Table~\ref{tab:chargetabley-1}, compatible with normal and
  inverted hierarchical neutrinos,
\item case 12c of Table~\ref{tab:chargetabley-1}, compatible with normal
  hierarchical neutrinos,
\item case 10b of Table~\ref{tab:chargetabley1}, compatible with degenerate
  neutrinos. 
\end{itemize}

One of the key outcomes of this paper results from the insistence of the
residual discrete $\ZR$ symmetry. By promoting the FN $\text{U(1)}$ to an
$R$-symmetry and by fixing the charge of the flavon field to $\qR[\phi]=-1$
we are left with the charge of the superspace coordinate $\theta$ as a free
parameter. However this parameter is fixed by the residual symmetry, which in
the case of a $\ZR$ symmetry results in $\qR[\theta]=\frac{2\mathbb{Z}+1}{4}$,
that is an odd-integer multiple of a quarter.

\section*{Acknowledgements}
\noindent
We would like to thank Felix Br\"ummer, Michael Ratz and Graham Ross for
helpful discussions.
The work of H.D. was supported through the SFB TR-33 `The Dark Universe'.
C.L. acknowledges support from the  EU ITN grants UNILHC 237920 and 
INVISIBLES 289442, and thanks the Physics Institute in Bonn for hospitality.




\newpage
\appendix
\section{Table of charge assignments}\label{app:tables}
In this appendix we present the resulting possible charge assignments for each of the parameter values in \cref{tab:collect-final}, where $y=-1,0,1$ is displayed in \cref{tab:chargetabley-1,tab:chargetabley0,tab:chargetabley1} respectively. In the case where multiple $x$ values are possible for a given case we use a, b, c and d after the case number, which is given in the first column of \cref{tab:collect-final}, such that the case is distinguishable. It should be noted once again that $y=0$ corresponds to the scenario where the CKM matrix is in the best agreement with experiment without relying on tuning through the order-one coupling coefficients.

\squeezetable 
\begin{table}[h!]
\renewcommand\arraystretch{1.8}
\begin{center}
\begin{tabular}{CCCCCCCCCCCCCCCCCCCCC}
\toprule
\# & \qR[\Hd] & \qR[\Hu] & \qR[\Q_1] & \qR[\Q_2] & \qR[\Q_3] & \qR[\U_1] & \qR[\U_2] & \qR[\U_3] & \qR[\D_1] & \qR[\D_2] & \qR[\D_3] & \qR[\Ld_1] & \qR[\Ld_2]& \qR[\Ld_3] & \qR[\E_1] & \qR[\E_2] & \qR[\E_3]\\[2.0pt] \midrule
%
 \text{1a} & -\frac{189}{170} & -\frac{321}{170} &
   \frac{1231}{340} & \frac{1231}{340} & \frac{551}{340} &
   \frac{1961}{340} & \frac{601}{340} & -\frac{79}{340} &
   \frac{337}{340} & -\frac{343}{340} & -\frac{343}{340} &
   \frac{47}{340} & \frac{47}{340} & -\frac{293}{340} &
   \frac{1861}{340} & \frac{841}{340} & \frac{501}{340} \\
 \text{1b} & -\frac{129}{190} & -\frac{441}{190} &
   \frac{4343}{1140} & \frac{4343}{1140} & \frac{2063}{1140} &
   \frac{6853}{1140} & \frac{2293}{1140} & \frac{13}{1140} &
   \frac{1561}{1140} & -\frac{719}{1140} & -\frac{719}{1140} &
   \frac{217}{380} & \frac{217}{380} & -\frac{163}{380} &
   \frac{2131}{380} & \frac{991}{380} & \frac{611}{380} \\
 \text{1c} & -\frac{53}{210} & -\frac{577}{210} &
   \frac{5041}{1260} & \frac{5041}{1260} & \frac{2521}{1260} &
   \frac{7871}{1260} & \frac{2831}{1260} & \frac{311}{1260} &
   \frac{2207}{1260} & -\frac{313}{1260} & -\frac{313}{1260} &
   \frac{419}{420} & \frac{419}{420} & -\frac{1}{420} &
   \frac{2417}{420} & \frac{1157}{420} & \frac{737}{420} \\
 \text{1d} & \frac{39}{230} & -\frac{729}{230} & \frac{1929}{460}
   & \frac{1929}{460} & \frac{1009}{460} & \frac{2979}{460} &
   \frac{1139}{460} & \frac{219}{460} & \frac{983}{460} &
   \frac{63}{460} & \frac{63}{460} & \frac{653}{460} &
   \frac{653}{460} & \frac{193}{460} & \frac{2719}{460} &
   \frac{1339}{460} & \frac{879}{460} \\ 
   %
 \text{2a} & -\frac{87}{190} & -\frac{483}{190} & \frac{1483}{380}
   & \frac{1483}{380} & \frac{723}{380} & \frac{2333}{380} &
   \frac{813}{380} & \frac{53}{380} & \frac{401}{380} &
   -\frac{359}{380} & -\frac{359}{380} & \frac{111}{380} &
   \frac{111}{380} & -\frac{269}{380} & \frac{2153}{380} &
   \frac{1013}{380} & \frac{633}{380} \\
 \text{2b} & -\frac{1}{30} & -\frac{89}{30} & \frac{737}{180} &
   \frac{737}{180} & \frac{377}{180} & \frac{1147}{180} &
   \frac{427}{180} & \frac{67}{180} & \frac{259}{180} &
   -\frac{101}{180} & -\frac{101}{180} & \frac{43}{60} &
   \frac{43}{60} & -\frac{17}{60} & \frac{349}{60} &
   \frac{169}{60} & \frac{109}{60} \\
 \text{2c} & \frac{89}{230} & -\frac{779}{230} & \frac{5917}{1380}
   & \frac{5917}{1380} & \frac{3157}{1380} & \frac{9107}{1380} &
   \frac{3587}{1380} & \frac{827}{1380} & \frac{2519}{1380} &
   -\frac{241}{1380} & -\frac{241}{1380} & \frac{523}{460} &
   \frac{523}{460} & \frac{63}{460} & \frac{2749}{460} &
   \frac{1369}{460} & \frac{909}{460} \\
      %
 \text{3a} & -\frac{3}{5} & -\frac{7}{5} & \frac{69}{20} &
   \frac{69}{20} & \frac{29}{20} & \frac{109}{20} & \frac{29}{20}
   & -\frac{11}{20} & \frac{13}{20} & -\frac{27}{20} &
   -\frac{27}{20} & -\frac{7}{20} & -\frac{27}{20} &
   -\frac{27}{20} & \frac{89}{20} & \frac{69}{20} & \frac{29}{20}
   \\
 \text{3b} & -\frac{13}{85} & -\frac{157}{85} & \frac{3707}{1020}
   & \frac{3707}{1020} & \frac{1667}{1020} & \frac{5827}{1020} &
   \frac{1747}{1020} & -\frac{293}{1020} & \frac{1039}{1020} &
   -\frac{1001}{1020} & -\frac{1001}{1020} & \frac{33}{340} &
   -\frac{307}{340} & -\frac{307}{340} & \frac{1549}{340} &
   \frac{1209}{340} & \frac{529}{340} \\
 \text{3c} & \frac{27}{95} & -\frac{217}{95} & \frac{4357}{1140} &
   \frac{4357}{1140} & \frac{2077}{1140} & \frac{6797}{1140} &
   \frac{2237}{1140} & -\frac{43}{1140} & \frac{1589}{1140} &
   -\frac{691}{1140} & -\frac{691}{1140} & \frac{203}{380} &
   -\frac{177}{380} & -\frac{177}{380} & \frac{1779}{380} &
   \frac{1399}{380} & \frac{639}{380} \\
 \text{3d} & \frac{5}{7} & -\frac{19}{7} & \frac{337}{84} &
   \frac{337}{84} & \frac{169}{84} & \frac{521}{84} &
   \frac{185}{84} & \frac{17}{84} & \frac{149}{84} &
   -\frac{19}{84} & -\frac{19}{84} & \frac{27}{28} & -\frac{1}{28}
   & -\frac{1}{28} & \frac{135}{28} & \frac{107}{28} &
   \frac{51}{28} \\
         %
\text{4a} & \frac{47}{95} & -\frac{237}{95} & \frac{1489}{380} & \frac{1489}{380} & \frac{729}{380} & \frac{2309}{380} & \frac{789}{380} & \frac{29}{380} &
   \frac{413}{380} & -\frac{347}{380} & -\frac{347}{380} & \frac{93}{380} & -\frac{287}{380} & -\frac{287}{380} & \frac{1809}{380} & \frac{1429}{380} & \frac{669}{380}
   \\
 \text{4b} & \frac{97}{105} & -\frac{307}{105} & \frac{5177}{1260} & \frac{5177}{1260} & \frac{2657}{1260} & \frac{7957}{1260} & \frac{2917}{1260} & \frac{397}{1260} &
   \frac{1849}{1260} & -\frac{671}{1260} & -\frac{671}{1260} & \frac{283}{420} & -\frac{137}{420} & -\frac{137}{420} & \frac{2059}{420} & \frac{1639}{420} &
   \frac{799}{420} \\
             %
 \text{5} & \frac{123}{110} & -\frac{13}{110} & \frac{2119}{660}
   & \frac{2119}{660} & \frac{799}{660} & \frac{3569}{660} &
   \frac{929}{660} & -\frac{391}{660} & \frac{113}{660} &
   -\frac{1207}{660} & -\frac{1207}{660} & -\frac{359}{220} &
   -\frac{359}{220} & -\frac{359}{220} & \frac{1323}{220} &
   \frac{663}{220} & \frac{223}{220} \\
                  %
 \text{6a} & \frac{29}{22} & -\frac{7}{22} & \frac{437}{132} &
   \frac{437}{132} & \frac{173}{132} & \frac{727}{132} &
   \frac{199}{132} & -\frac{65}{132} & -\frac{17}{132} &
   -\frac{281}{132} & -\frac{281}{132} & -\frac{85}{44} &
   -\frac{85}{44} & -\frac{85}{44} & \frac{269}{44} &
   \frac{137}{44} & \frac{49}{44} \\
 \text{6b} & \frac{229}{130} & -\frac{99}{130} & \frac{909}{260} &
   \frac{909}{260} & \frac{389}{260} & \frac{1499}{260} &
   \frac{459}{260} & -\frac{61}{260} & \frac{63}{260} &
   -\frac{457}{260} & -\frac{457}{260} & -\frac{387}{260} &
   -\frac{387}{260} & -\frac{387}{260} & \frac{1619}{260} &
   \frac{839}{260} & \frac{319}{260} \\
 \text{6c} & \frac{329}{150} & -\frac{179}{150} & \frac{3317}{900}
   & \frac{3317}{900} & \frac{1517}{900} & \frac{5407}{900} &
   \frac{1807}{900} & \frac{7}{900} & \frac{559}{900} &
   -\frac{1241}{900} & -\frac{1241}{900} & -\frac{317}{300} &
   -\frac{317}{300} & -\frac{317}{300} & \frac{1909}{300} &
   \frac{1009}{300} & \frac{409}{300} \\
 \text{6d} & \frac{89}{34} & -\frac{55}{34} & \frac{791}{204} &
   \frac{791}{204} & \frac{383}{204} & \frac{1273}{204} &
   \frac{457}{204} & \frac{49}{204} & \frac{205}{204} &
   -\frac{203}{204} & -\frac{203}{204} & -\frac{43}{68} &
   -\frac{43}{68} & -\frac{43}{68} & \frac{443}{68} &
   \frac{239}{68} & \frac{103}{68} \\
 \text{7a} & \frac{29}{22} & -\frac{7}{22} & \frac{437}{132} &
   \frac{437}{132} & \frac{173}{132} & \frac{727}{132} &
   \frac{199}{132} & -\frac{65}{132} & -\frac{17}{132} &
   -\frac{281}{132} & -\frac{281}{132} & -\frac{85}{44} &
   -\frac{85}{44} & -\frac{85}{44} & \frac{269}{44} &
   \frac{137}{44} & \frac{49}{44} \\
 \text{7b} & \frac{229}{130} & -\frac{99}{130} & \frac{909}{260} &
   \frac{909}{260} & \frac{389}{260} & \frac{1499}{260} &
   \frac{459}{260} & -\frac{61}{260} & \frac{63}{260} &
   -\frac{457}{260} & -\frac{457}{260} & -\frac{387}{260} &
   -\frac{387}{260} & -\frac{387}{260} & \frac{1619}{260} &
   \frac{839}{260} & \frac{319}{260} \\
   %
 \text{8a} & \frac{167}{110} & -\frac{57}{110} & \frac{2251}{660}
   & \frac{2251}{660} & \frac{931}{660} & \frac{3701}{660} &
   \frac{1061}{660} & -\frac{259}{660} & -\frac{283}{660} &
   -\frac{1603}{660} & -\frac{1603}{660} & -\frac{491}{220} &
   -\frac{491}{220} & -\frac{491}{220} & \frac{1367}{220} &
   \frac{707}{220} & \frac{267}{220} \\
 \text{8b} & \frac{51}{26} & -\frac{25}{26} & \frac{187}{52} &
   \frac{187}{52} & \frac{83}{52} & \frac{305}{52} & \frac{97}{52}
   & -\frac{7}{52} & -\frac{3}{52} & -\frac{107}{52} &
   -\frac{107}{52} & -\frac{93}{52} & -\frac{93}{52} &
   -\frac{93}{52} & \frac{329}{52} & \frac{173}{52} &
   \frac{69}{52} \\
 \text{8c} & \frac{359}{150} & -\frac{209}{150} & \frac{3407}{900}
   & \frac{3407}{900} & \frac{1607}{900} & \frac{5497}{900} &
   \frac{1897}{900} & \frac{97}{900} & \frac{289}{900} &
   -\frac{1511}{900} & -\frac{1511}{900} & -\frac{407}{300} &
   -\frac{407}{300} & -\frac{407}{300} & \frac{1939}{300} &
   \frac{1039}{300} & \frac{439}{300} \\
 \text{8d} & \frac{479}{170} & -\frac{309}{170} &
   \frac{4057}{1020} & \frac{4057}{1020} & \frac{2017}{1020} &
   \frac{6467}{1020} & \frac{2387}{1020} & \frac{347}{1020} &
   \frac{719}{1020} & -\frac{1321}{1020} & -\frac{1321}{1020} &
   -\frac{317}{340} & -\frac{317}{340} & -\frac{317}{340} &
   \frac{2249}{340} & \frac{1229}{340} & \frac{549}{340} \\
   %
 \text{9a} & \frac{189}{110} & -\frac{79}{110} & \frac{2317}{660}
   & \frac{2317}{660} & \frac{997}{660} & \frac{3767}{660} &
   \frac{1127}{660} & -\frac{193}{660} & -\frac{481}{660} &
   -\frac{1801}{660} & -\frac{1801}{660} & -\frac{557}{220} &
   -\frac{557}{220} & -\frac{557}{220} & \frac{1389}{220} &
   \frac{729}{220} & \frac{289}{220} \\
 \text{9b} & \frac{281}{130} & -\frac{151}{130} & \frac{961}{260}
   & \frac{961}{260} & \frac{441}{260} & \frac{1551}{260} &
   \frac{511}{260} & -\frac{9}{260} & -\frac{93}{260} &
   -\frac{613}{260} & -\frac{613}{260} & -\frac{543}{260} &
   -\frac{543}{260} & -\frac{543}{260} & \frac{1671}{260} &
   \frac{891}{260} & \frac{371}{260} \\
 \text{9c} & \frac{389}{150} & -\frac{239}{150} &
   \frac{3497}{900} & \frac{3497}{900} & \frac{1697}{900} &
   \frac{5587}{900} & \frac{1987}{900} & \frac{187}{900} &
   \frac{19}{900} & -\frac{1781}{900} & -\frac{1781}{900} &
   -\frac{497}{300} & -\frac{497}{300} & -\frac{497}{300} &
   \frac{1969}{300} & \frac{1069}{300} & \frac{469}{300} \\
 \text{9d} & \frac{513}{170} & -\frac{343}{170} &
   \frac{4159}{1020} & \frac{4159}{1020} & \frac{2119}{1020} &
   \frac{6569}{1020} & \frac{2489}{1020} & \frac{449}{1020} &
   \frac{413}{1020} & -\frac{1627}{1020} & -\frac{1627}{1020} &
   -\frac{419}{340} & -\frac{419}{340} & -\frac{419}{340} &
   \frac{2283}{340} & \frac{1263}{340} & \frac{583}{340} \\
   %
 \text{10a} & \frac{419}{150} & -\frac{269}{150} &
   \frac{3587}{900} & \frac{3587}{900} & \frac{1787}{900} &
   \frac{5677}{900} & \frac{2077}{900} & \frac{277}{900} &
   -\frac{251}{900} & -\frac{2051}{900} & -\frac{2051}{900} &
   -\frac{587}{300} & -\frac{587}{300} & -\frac{587}{300} &
   \frac{1999}{300} & \frac{1099}{300} & \frac{499}{300} \\
 \text{10b} & \frac{547}{170} & -\frac{377}{170} &
   \frac{4261}{1020} & \frac{4261}{1020} & \frac{2221}{1020} &
   \frac{6671}{1020} & \frac{2591}{1020} & \frac{551}{1020} &
   \frac{107}{1020} & -\frac{1933}{1020} & -\frac{1933}{1020} &
   -\frac{521}{340} & -\frac{521}{340} & -\frac{521}{340} &
   \frac{2317}{340} & \frac{1297}{340} & \frac{617}{340} \\
   %
 \text{11} & \frac{81}{55} & \frac{29}{55} & \frac{2041}{660} &
   \frac{2041}{660} & \frac{721}{660} & \frac{3221}{660} &
   \frac{581}{660} & -\frac{739}{660} & \frac{617}{660} &
   -\frac{703}{660} & -\frac{703}{660} & -\frac{281}{220} &
   -\frac{281}{220} & -\frac{501}{220} & \frac{1167}{220} &
   \frac{727}{220} & \frac{507}{220} \\
   %
 \text{12a} & \frac{91}{55} & \frac{19}{55} & \frac{2111}{660} &
   \frac{2111}{660} & \frac{791}{660} & \frac{3271}{660} &
   \frac{631}{660} & -\frac{689}{660} & \frac{427}{660} &
   -\frac{893}{660} & -\frac{893}{660} & -\frac{351}{220} &
   -\frac{351}{220} & -\frac{571}{220} & \frac{1197}{220} &
   \frac{757}{220} & \frac{537}{220} \\
 \text{12b} & \frac{139}{65} & -\frac{9}{65} & \frac{2629}{780} &
   \frac{2629}{780} & \frac{1069}{780} & \frac{4109}{780} &
   \frac{989}{780} & -\frac{571}{780} & \frac{773}{780} &
   -\frac{787}{780} & -\frac{787}{780} & -\frac{289}{260} &
   -\frac{289}{260} & -\frac{549}{260} & \frac{1423}{260} &
   \frac{903}{260} & \frac{643}{260} \\
 \text{12c} & \frac{13}{5} & -\frac{3}{5} & \frac{71}{20} &
   \frac{71}{20} & \frac{31}{20} & \frac{111}{20} & \frac{31}{20}
   & -\frac{9}{20} & \frac{27}{20} & -\frac{13}{20} &
   -\frac{13}{20} & -\frac{13}{20} & -\frac{13}{20} &
   -\frac{33}{20} & \frac{111}{20} & \frac{71}{20} & \frac{51}{20}
   \\ \bottomrule
\end{tabular}
\end{center}
\caption{Charge assignments of the fields under the $\text{U}(1)_R$ symmetry given the parameter values shown in Table~\ref{tab:collect-final} for the scenario where $y=-1$.}
\label{tab:chargetabley-1}
\end{table}

\squeezetable 
\begin{table}[h!]
\renewcommand\arraystretch{1.8}
\begin{center}
\begin{tabular}{CCCCCCCCCCCCCCCCCCCCC}
\toprule
\# & \qR[\Hd] & \qR[\Hu] & \qR[\Q_1] & \qR[\Q_2] & \qR[\Q_3] & \qR[\U_1] & \qR[\U_2] & \qR[\U_3] & \qR[\D_1] & \qR[\D_2] & \qR[\D_3] & \qR[\Ld_1] & \qR[\Ld_2]& \qR[\Ld_3] & \qR[\E_1] & \qR[\E_2] & \qR[\E_3]\\[2.0pt] \midrule
%
 \text{1a} & -\frac{33}{34} & -\frac{69}{34} & \frac{865}{204} & \frac{661}{204} & \frac{253}{204} & \frac{1079}{204} & \frac{467}{204} & \frac{59}{204} &
   \frac{47}{204} & -\frac{157}{204} & -\frac{157}{204} & \frac{19}{68} & \frac{19}{68} & -\frac{49}{68} & \frac{353}{68} & \frac{149}{68} & \frac{81}{68} \\
 \text{1b} & -\frac{21}{38} & -\frac{93}{38} & \frac{337}{76} & \frac{261}{76} & \frac{109}{76} & \frac{419}{76} & \frac{191}{76} & \frac{39}{76} & \frac{47}{76} &
   -\frac{29}{76} & -\frac{29}{76} & \frac{53}{76} & \frac{53}{76} & -\frac{23}{76} & \frac{407}{76} & \frac{179}{76} & \frac{103}{76} \\
 \text{1c} & -\frac{29}{210} & -\frac{601}{210} & \frac{5833}{1260} & \frac{4573}{1260} & \frac{2053}{1260} & \frac{7223}{1260} & \frac{3443}{1260} & \frac{923}{1260}
   & \frac{1271}{1260} & \frac{11}{1260} & \frac{11}{1260} & \frac{467}{420} & \frac{467}{420} & \frac{47}{420} & \frac{2321}{420} & \frac{1061}{420} &
   \frac{641}{420} \\
 \text{1d} & \frac{63}{230} & -\frac{753}{230} & \frac{6659}{1380} & \frac{5279}{1380} & \frac{2519}{1380} & \frac{8209}{1380} & \frac{4069}{1380} & \frac{1309}{1380}
   & \frac{1933}{1380} & \frac{553}{1380} & \frac{553}{1380} & \frac{701}{460} & \frac{701}{460} & \frac{241}{460} & \frac{2623}{460} & \frac{1243}{460} &
   \frac{783}{460} \\
%
 \text{2a} & -\frac{63}{190} & -\frac{507}{190} &
   \frac{5161}{1140} & \frac{4021}{1140} & \frac{1741}{1140} &
   \frac{6431}{1140} & \frac{3011}{1140} & \frac{731}{1140} &
   \frac{347}{1140} & -\frac{793}{1140} & -\frac{793}{1140} &
   \frac{159}{380} & \frac{159}{380} & -\frac{221}{380} &
   \frac{2057}{380} & \frac{917}{380} & \frac{537}{380} \\
 \text{2b} & \frac{17}{210} & -\frac{647}{210} & \frac{5951}{1260}
   & \frac{4691}{1260} & \frac{2171}{1260} & \frac{7381}{1260} &
   \frac{3601}{1260} & \frac{1081}{1260} & \frac{877}{1260} &
   -\frac{383}{1260} & -\frac{383}{1260} & \frac{349}{420} &
   \frac{349}{420} & -\frac{71}{420} & \frac{2347}{420} &
   \frac{1087}{420} & \frac{667}{420} \\
 \text{2c} & \frac{113}{230} & -\frac{803}{230} & \frac{2263}{460}
   & \frac{1803}{460} & \frac{883}{460} & \frac{2793}{460} &
   \frac{1413}{460} & \frac{493}{460} & \frac{501}{460} &
   \frac{41}{460} & \frac{41}{460} & \frac{571}{460} &
   \frac{571}{460} & \frac{111}{460} & \frac{2653}{460} &
   \frac{1273}{460} & \frac{813}{460} \\
   %
\text{3a} & -\frac{11}{25} & -\frac{39}{25} & \frac{1219}{300} &
   \frac{919}{300} & \frac{319}{300} & \frac{1499}{300} &
   \frac{599}{300} & -\frac{1}{300} & -\frac{37}{300} &
   -\frac{337}{300} & -\frac{337}{300} & -\frac{19}{100} &
   -\frac{119}{100} & -\frac{119}{100} & \frac{413}{100} &
   \frac{313}{100} & \frac{113}{100} \\
 \text{3b} & -\frac{1}{85} & -\frac{169}{85} & \frac{4339}{1020} &
   \frac{3319}{1020} & \frac{1279}{1020} & \frac{5339}{1020} &
   \frac{2279}{1020} & \frac{239}{1020} & \frac{263}{1020} &
   -\frac{757}{1020} & -\frac{757}{1020} & \frac{81}{340} &
   -\frac{259}{340} & -\frac{259}{340} & \frac{1453}{340} &
   \frac{1113}{340} & \frac{433}{340} \\
 \text{3c} & \frac{39}{95} & -\frac{229}{95} & \frac{5069}{1140} &
   \frac{3929}{1140} & \frac{1649}{1140} & \frac{6229}{1140} &
   \frac{2809}{1140} & \frac{529}{1140} & \frac{733}{1140} &
   -\frac{407}{1140} & -\frac{407}{1140} & \frac{251}{380} &
   -\frac{129}{380} & -\frac{129}{380} & \frac{1683}{380} &
   \frac{1303}{380} & \frac{543}{380} \\
 \text{3d} & \frac{29}{35} & -\frac{99}{35} & \frac{1949}{420} &
   \frac{1529}{420} & \frac{689}{420} & \frac{2389}{420} &
   \frac{1129}{420} & \frac{289}{420} & \frac{433}{420} &
   \frac{13}{420} & \frac{13}{420} & \frac{151}{140} &
   \frac{11}{140} & \frac{11}{140} & \frac{643}{140} &
   \frac{503}{140} & \frac{223}{140} \\
      %
\text{4a} & \frac{59}{95} & -\frac{249}{95} & \frac{5179}{1140} & \frac{4039}{1140} & \frac{1759}{1140} & \frac{6359}{1140} & \frac{2939}{1140} & \frac{659}{1140} &
   \frac{383}{1140} & -\frac{757}{1140} & -\frac{757}{1140} & \frac{141}{380} & -\frac{239}{380} & -\frac{239}{380} & \frac{1713}{380} & \frac{1333}{380} &
   \frac{573}{380} \\
 \text{4b} & \frac{109}{105} & -\frac{319}{105} & \frac{5969}{1260} & \frac{4709}{1260} & \frac{2189}{1260} & \frac{7309}{1260} & \frac{3529}{1260} & \frac{1009}{1260}
   & \frac{913}{1260} & -\frac{347}{1260} & -\frac{347}{1260} & \frac{331}{420} & -\frac{89}{420} & -\frac{89}{420} & \frac{1963}{420} & \frac{1543}{420} &
   \frac{703}{420} \\
                %
 \text{5} & \frac{147}{110} & -\frac{37}{110} & \frac{837}{220} &
   \frac{617}{220} & \frac{177}{220} & \frac{1107}{220} &
   \frac{447}{220} & \frac{7}{220} & -\frac{141}{220} &
   -\frac{361}{220} & -\frac{361}{220} & -\frac{311}{220} &
   -\frac{311}{220} & -\frac{311}{220} & \frac{1227}{220} &
   \frac{567}{220} & \frac{127}{220} \\
                %
 \text{6a} & \frac{169}{110} & -\frac{59}{110} & \frac{859}{220} &
   \frac{639}{220} & \frac{199}{220} & \frac{1129}{220} &
   \frac{469}{220} & \frac{29}{220} & -\frac{207}{220} &
   -\frac{427}{220} & -\frac{427}{220} & -\frac{377}{220} &
   -\frac{377}{220} & -\frac{377}{220} & \frac{1249}{220} &
   \frac{589}{220} & \frac{149}{220} \\
 \text{6b} & \frac{253}{130} & -\frac{123}{130} & \frac{3199}{780}
   & \frac{2419}{780} & \frac{859}{780} & \frac{4169}{780} &
   \frac{1829}{780} & \frac{269}{780} & -\frac{427}{780} &
   -\frac{1207}{780} & -\frac{1207}{780} & -\frac{339}{260} &
   -\frac{339}{260} & -\frac{339}{260} & \frac{1523}{260} &
   \frac{743}{260} & \frac{223}{260} \\
 \text{6c} & \frac{353}{150} & -\frac{203}{150} & \frac{3869}{900}
   & \frac{2969}{900} & \frac{1169}{900} & \frac{4999}{900} &
   \frac{2299}{900} & \frac{499}{900} & -\frac{137}{900} &
   -\frac{1037}{900} & -\frac{1037}{900} & -\frac{269}{300} &
   -\frac{269}{300} & -\frac{269}{300} & \frac{1813}{300} &
   \frac{913}{300} & \frac{313}{300} \\
 \text{6d} & \frac{469}{170} & -\frac{299}{170} & \frac{1529}{340}
   & \frac{1189}{340} & \frac{509}{340} & \frac{1959}{340} &
   \frac{939}{340} & \frac{259}{340} & \frac{83}{340} &
   -\frac{257}{340} & -\frac{257}{340} & -\frac{167}{340} &
   -\frac{167}{340} & -\frac{167}{340} & \frac{2119}{340} &
   \frac{1099}{340} & \frac{419}{340} \\
      %
 \text{7a} & \frac{169}{110} & -\frac{59}{110} & \frac{859}{220} &
   \frac{639}{220} & \frac{199}{220} & \frac{1129}{220} &
   \frac{469}{220} & \frac{29}{220} & -\frac{207}{220} &
   -\frac{427}{220} & -\frac{427}{220} & -\frac{377}{220} &
   -\frac{377}{220} & -\frac{377}{220} & \frac{1249}{220} &
   \frac{589}{220} & \frac{149}{220} \\
 \text{7b} & \frac{253}{130} & -\frac{123}{130} & \frac{3199}{780}
   & \frac{2419}{780} & \frac{859}{780} & \frac{4169}{780} &
   \frac{1829}{780} & \frac{269}{780} & -\frac{427}{780} &
   -\frac{1207}{780} & -\frac{1207}{780} & -\frac{339}{260} &
   -\frac{339}{260} & -\frac{339}{260} & \frac{1523}{260} &
   \frac{743}{260} & \frac{223}{260} \\
   %
 \text{8a} & \frac{191}{110} & -\frac{81}{110} & \frac{881}{220} &
   \frac{661}{220} & \frac{221}{220} & \frac{1151}{220} &
   \frac{491}{220} & \frac{51}{220} & -\frac{273}{220} &
   -\frac{493}{220} & -\frac{493}{220} & -\frac{443}{220} &
   -\frac{443}{220} & -\frac{443}{220} & \frac{1271}{220} &
   \frac{611}{220} & \frac{171}{220} \\
 \text{8b} & \frac{279}{130} & -\frac{149}{130} & \frac{3277}{780}
   & \frac{2497}{780} & \frac{937}{780} & \frac{4247}{780} &
   \frac{1907}{780} & \frac{347}{780} & -\frac{661}{780} &
   -\frac{1441}{780} & -\frac{1441}{780} & -\frac{417}{260} &
   -\frac{417}{260} & -\frac{417}{260} & \frac{1549}{260} &
   \frac{769}{260} & \frac{249}{260} \\
 \text{8c} & \frac{383}{150} & -\frac{233}{150} & \frac{3959}{900}
   & \frac{3059}{900} & \frac{1259}{900} & \frac{5089}{900} &
   \frac{2389}{900} & \frac{589}{900} & -\frac{407}{900} &
   -\frac{1307}{900} & -\frac{1307}{900} & -\frac{359}{300} &
   -\frac{359}{300} & -\frac{359}{300} & \frac{1843}{300} &
   \frac{943}{300} & \frac{343}{300} \\
 \text{8d} & \frac{503}{170} & -\frac{333}{170} & \frac{1563}{340}
   & \frac{1223}{340} & \frac{543}{340} & \frac{1993}{340} &
   \frac{973}{340} & \frac{293}{340} & -\frac{19}{340} &
   -\frac{359}{340} & -\frac{359}{340} & -\frac{269}{340} &
   -\frac{269}{340} & -\frac{269}{340} & \frac{2153}{340} &
   \frac{1133}{340} & \frac{453}{340} \\
   %
 \text{9a} & \frac{213}{110} & -\frac{103}{110} & \frac{903}{220}
   & \frac{683}{220} & \frac{243}{220} & \frac{1173}{220} &
   \frac{513}{220} & \frac{73}{220} & -\frac{339}{220} &
   -\frac{559}{220} & -\frac{559}{220} & -\frac{509}{220} &
   -\frac{509}{220} & -\frac{509}{220} & \frac{1293}{220} &
   \frac{633}{220} & \frac{193}{220} \\
 \text{9b} & \frac{61}{26} & -\frac{35}{26} & \frac{671}{156} &
   \frac{515}{156} & \frac{203}{156} & \frac{865}{156} &
   \frac{397}{156} & \frac{85}{156} & -\frac{179}{156} &
   -\frac{335}{156} & -\frac{335}{156} & -\frac{99}{52} &
   -\frac{99}{52} & -\frac{99}{52} & \frac{315}{52} &
   \frac{159}{52} & \frac{55}{52} \\
 \text{9c} & \frac{413}{150} & -\frac{263}{150} &
   \frac{4049}{900} & \frac{3149}{900} & \frac{1349}{900} &
   \frac{5179}{900} & \frac{2479}{900} & \frac{679}{900} &
   -\frac{677}{900} & -\frac{1577}{900} & -\frac{1577}{900} &
   -\frac{449}{300} & -\frac{449}{300} & -\frac{449}{300} &
   \frac{1873}{300} & \frac{973}{300} & \frac{373}{300} \\
 \text{9d} & \frac{537}{170} & -\frac{367}{170} &
   \frac{1597}{340} & \frac{1257}{340} & \frac{577}{340} &
   \frac{2027}{340} & \frac{1007}{340} & \frac{327}{340} &
   -\frac{121}{340} & -\frac{461}{340} & -\frac{461}{340} &
   -\frac{371}{340} & -\frac{371}{340} & -\frac{371}{340} &
   \frac{2187}{340} & \frac{1167}{340} & \frac{487}{340} \\
   %
 \text{10a} & \frac{443}{150} & -\frac{293}{150} &
   \frac{4139}{900} & \frac{3239}{900} & \frac{1439}{900} &
   \frac{5269}{900} & \frac{2569}{900} & \frac{769}{900} &
   -\frac{947}{900} & -\frac{1847}{900} & -\frac{1847}{900} &
   -\frac{539}{300} & -\frac{539}{300} & -\frac{539}{300} &
   \frac{1903}{300} & \frac{1003}{300} & \frac{403}{300} \\
 \text{10b} & \frac{571}{170} & -\frac{401}{170} &
   \frac{1631}{340} & \frac{1291}{340} & \frac{611}{340} &
   \frac{2061}{340} & \frac{1041}{340} & \frac{361}{340} &
   -\frac{223}{340} & -\frac{563}{340} & -\frac{563}{340} &
   -\frac{473}{340} & -\frac{473}{340} & -\frac{473}{340} &
   \frac{2221}{340} & \frac{1201}{340} & \frac{521}{340} \\
   %
 \text{11} & \frac{93}{55} & \frac{17}{55} & \frac{811}{220} &
   \frac{591}{220} & \frac{151}{220} & \frac{991}{220} &
   \frac{331}{220} & -\frac{109}{220} & \frac{27}{220} &
   -\frac{193}{220} & -\frac{193}{220} & -\frac{233}{220} &
   -\frac{233}{220} & -\frac{453}{220} & \frac{1071}{220} &
   \frac{631}{220} & \frac{411}{220} \\
   %
 \text{12a} & \frac{103}{55} & \frac{7}{55} & \frac{2503}{660} &
   \frac{1843}{660} & \frac{523}{660} & \frac{3023}{660} &
   \frac{1043}{660} & -\frac{277}{660} & -\frac{109}{660} &
   -\frac{769}{660} & -\frac{769}{660} & -\frac{303}{220} &
   -\frac{303}{220} & -\frac{523}{220} & \frac{1101}{220} &
   \frac{661}{220} & \frac{441}{220} \\
 \text{12b} & \frac{151}{65} & -\frac{21}{65} & \frac{3101}{780} &
   \frac{2321}{780} & \frac{761}{780} & \frac{3781}{780} &
   \frac{1441}{780} & -\frac{119}{780} & \frac{157}{780} &
   -\frac{623}{780} & -\frac{623}{780} & -\frac{241}{260} &
   -\frac{241}{260} & -\frac{501}{260} & \frac{1327}{260} &
   \frac{807}{260} & \frac{547}{260} \\
 \text{12c} & \frac{69}{25} & -\frac{19}{25} & \frac{1249}{300} &
   \frac{949}{300} & \frac{349}{300} & \frac{1529}{300} &
   \frac{629}{300} & \frac{29}{300} & \frac{173}{300} &
   -\frac{127}{300} & -\frac{127}{300} & -\frac{49}{100} &
   -\frac{49}{100} & -\frac{149}{100} & \frac{523}{100} &
   \frac{323}{100} & \frac{223}{100} \\
   \bottomrule
\end{tabular}
\end{center}
\caption{Charge assignments of the fields under the $\text{U}(1)_R$ symmetry given the parameter values shown in Table~\ref{tab:collect-final} for the scenario where $y=0$.}
\label{tab:chargetabley0}
\end{table}

\begin{table}[h!]
\renewcommand\arraystretch{1.8}
\begin{center}
\begin{tabular}{CCCCCCCCCCCCCCCCCCCCC}
\toprule
\# & \qR[\Hd] & \qR[\Hu] & \qR[\Q_1] & \qR[\Q_2] & \qR[\Q_3] & \qR[\U_1] & \qR[\U_2] & \qR[\U_3] & \qR[\D_1] & \qR[\D_2] & \qR[\D_3] & \qR[\Ld_1] & \qR[\Ld_2]& \qR[\Ld_3] & \qR[\E_1] & \qR[\E_2] & \qR[\E_3]\\[2.0pt] \midrule
%
 \text{1a} & -\frac{141}{170} & -\frac{369}{170} &
   \frac{4957}{1020} & \frac{2917}{1020} & \frac{877}{1020} &
   \frac{4907}{1020} & \frac{2867}{1020} & \frac{827}{1020} &
   -\frac{541}{1020} & -\frac{541}{1020} & -\frac{541}{1020} &
   \frac{143}{340} & \frac{143}{340} & -\frac{197}{340} &
   \frac{1669}{340} & \frac{649}{340} & \frac{309}{340} \\
 \text{1b} & -\frac{81}{190} & -\frac{489}{190} &
   \frac{5767}{1140} & \frac{3487}{1140} & \frac{1207}{1140} &
   \frac{5717}{1140} & \frac{3437}{1140} & \frac{1157}{1140} &
   -\frac{151}{1140} & -\frac{151}{1140} & -\frac{151}{1140} &
   \frac{313}{380} & \frac{313}{380} & -\frac{67}{380} &
   \frac{1939}{380} & \frac{799}{380} & \frac{419}{380} \\
 \text{1c} & -\frac{1}{42} & -\frac{125}{42} & \frac{1325}{252} &
   \frac{821}{252} & \frac{317}{252} & \frac{1315}{252} &
   \frac{811}{252} & \frac{307}{252} & \frac{67}{252} &
   \frac{67}{252} & \frac{67}{252} & \frac{103}{84} &
   \frac{103}{84} & \frac{19}{84} & \frac{445}{84} &
   \frac{193}{84} & \frac{109}{84} \\
 \text{1d} & \frac{87}{230} & -\frac{777}{230} & \frac{7531}{1380}
   & \frac{4771}{1380} & \frac{2011}{1380} & \frac{7481}{1380} &
   \frac{4721}{1380} & \frac{1961}{1380} & \frac{917}{1380} &
   \frac{917}{1380} & \frac{917}{1380} & \frac{749}{460} &
   \frac{749}{460} & \frac{289}{460} & \frac{2527}{460} &
   \frac{1147}{460} & \frac{687}{460} \\
   %
 \text{2a} & -\frac{39}{190} & -\frac{531}{190} &
   \frac{5873}{1140} & \frac{3593}{1140} & \frac{1313}{1140} &
   \frac{5863}{1140} & \frac{3583}{1140} & \frac{1303}{1140} &
   -\frac{509}{1140} & -\frac{509}{1140} & -\frac{509}{1140} &
   \frac{207}{380} & \frac{207}{380} & -\frac{173}{380} &
   \frac{1961}{380} & \frac{821}{380} & \frac{441}{380} \\
 \text{2b} & \frac{41}{210} & -\frac{671}{210} & \frac{6743}{1260}
   & \frac{4223}{1260} & \frac{1703}{1260} & \frac{6733}{1260} &
   \frac{4213}{1260} & \frac{1693}{1260} & -\frac{59}{1260} &
   -\frac{59}{1260} & -\frac{59}{1260} & \frac{397}{420} &
   \frac{397}{420} & -\frac{23}{420} & \frac{2251}{420} &
   \frac{991}{420} & \frac{571}{420} \\
 \text{2c} & \frac{137}{230} & -\frac{827}{230} &
   \frac{7661}{1380} & \frac{4901}{1380} & \frac{2141}{1380} &
   \frac{7651}{1380} & \frac{4891}{1380} & \frac{2131}{1380} &
   \frac{487}{1380} & \frac{487}{1380} & \frac{487}{1380} &
   \frac{619}{460} & \frac{619}{460} & \frac{159}{460} &
   \frac{2557}{460} & \frac{1177}{460} & \frac{717}{460} \\
      %
 \text{3a} & -\frac{7}{25} & -\frac{43}{25} & \frac{1403}{300} &
   \frac{803}{300} & \frac{203}{300} & \frac{1363}{300} &
   \frac{763}{300} & \frac{163}{300} & -\frac{269}{300} &
   -\frac{269}{300} & -\frac{269}{300} & -\frac{3}{100} &
   -\frac{103}{100} & -\frac{103}{100} & \frac{381}{100} &
   \frac{281}{100} & \frac{81}{100} \\
 \text{3b} & \frac{11}{85} & -\frac{181}{85} & \frac{1657}{340} &
   \frac{977}{340} & \frac{297}{340} & \frac{1617}{340} &
   \frac{937}{340} & \frac{257}{340} & -\frac{171}{340} &
   -\frac{171}{340} & -\frac{171}{340} & \frac{129}{340} &
   -\frac{211}{340} & -\frac{211}{340} & \frac{1357}{340} &
   \frac{1017}{340} & \frac{337}{340} \\
 \text{3c} & \frac{51}{95} & -\frac{241}{95} & \frac{1927}{380} &
   \frac{1167}{380} & \frac{407}{380} & \frac{1887}{380} &
   \frac{1127}{380} & \frac{367}{380} & -\frac{41}{380} &
   -\frac{41}{380} & -\frac{41}{380} & \frac{299}{380} &
   -\frac{81}{380} & -\frac{81}{380} & \frac{1587}{380} &
   \frac{1207}{380} & \frac{447}{380} \\
 \text{3d} & \frac{33}{35} & -\frac{103}{35} & \frac{2213}{420} &
   \frac{1373}{420} & \frac{533}{420} & \frac{2173}{420} &
   \frac{1333}{420} & \frac{493}{420} & \frac{121}{420} &
   \frac{121}{420} & \frac{121}{420} & \frac{167}{140} &
   \frac{27}{140} & \frac{27}{140} & \frac{611}{140} &
   \frac{471}{140} & \frac{191}{140} \\
         %
 \text{4a} & \frac{71}{95} & -\frac{261}{95} & \frac{5891}{1140} & \frac{3611}{1140} & \frac{1331}{1140} & \frac{5791}{1140} & \frac{3511}{1140} & \frac{1231}{1140} &
   -\frac{473}{1140} & -\frac{473}{1140} & -\frac{473}{1140} & \frac{189}{380} & -\frac{191}{380} & -\frac{191}{380} & \frac{1617}{380} & \frac{1237}{380} &
   \frac{477}{380} \\
 \text{4b} & \frac{121}{105} & -\frac{331}{105} & \frac{6761}{1260} & \frac{4241}{1260} & \frac{1721}{1260} & \frac{6661}{1260} & \frac{4141}{1260} & \frac{1621}{1260}
   & -\frac{23}{1260} & -\frac{23}{1260} & -\frac{23}{1260} & \frac{379}{420} & -\frac{41}{420} & -\frac{41}{420} & \frac{1867}{420} & \frac{1447}{420} &
   \frac{607}{420} \\
          %
    \text{5} & \frac{171}{110} & -\frac{61}{110} & \frac{2903}{660}
   & \frac{1583}{660} & \frac{263}{660} & \frac{3073}{660} &
   \frac{1753}{660} & \frac{433}{660} & -\frac{959}{660} &
   -\frac{959}{660} & -\frac{959}{660} & -\frac{263}{220} &
   -\frac{263}{220} & -\frac{263}{220} & \frac{1131}{220} &
   \frac{471}{220} & \frac{31}{220} \\
                  %
\text{6a} & \frac{193}{110} & -\frac{83}{110} & \frac{2969}{660}
   & \frac{1649}{660} & \frac{329}{660} & \frac{3139}{660} &
   \frac{1819}{660} & \frac{499}{660} & -\frac{1157}{660} &
   -\frac{1157}{660} & -\frac{1157}{660} & -\frac{329}{220} &
   -\frac{329}{220} & -\frac{329}{220} & \frac{1153}{220} &
   \frac{493}{220} & \frac{53}{220} \\
 \text{6b} & \frac{277}{130} & -\frac{147}{130} & \frac{3671}{780}
   & \frac{2111}{780} & \frac{551}{780} & \frac{3841}{780} &
   \frac{2281}{780} & \frac{721}{780} & -\frac{1043}{780} &
   -\frac{1043}{780} & -\frac{1043}{780} & -\frac{291}{260} &
   -\frac{291}{260} & -\frac{291}{260} & \frac{1427}{260} &
   \frac{647}{260} & \frac{127}{260} \\
 \text{6c} & \frac{377}{150} & -\frac{227}{150} & \frac{4421}{900}
   & \frac{2621}{900} & \frac{821}{900} & \frac{4591}{900} &
   \frac{2791}{900} & \frac{991}{900} & -\frac{833}{900} &
   -\frac{833}{900} & -\frac{833}{900} & -\frac{221}{300} &
   -\frac{221}{300} & -\frac{221}{300} & \frac{1717}{300} &
   \frac{817}{300} & \frac{217}{300} \\
 \text{6d} & \frac{29}{10} & -\frac{19}{10} & \frac{307}{60} &
   \frac{187}{60} & \frac{67}{60} & \frac{317}{60} &
   \frac{197}{60} & \frac{77}{60} & -\frac{31}{60} &
   -\frac{31}{60} & -\frac{31}{60} & -\frac{7}{20} & -\frac{7}{20}
   & -\frac{7}{20} & \frac{119}{20} & \frac{59}{20} &
   \frac{19}{20} \\
         %
 \text{7a} & \frac{193}{110} & -\frac{83}{110} & \frac{2969}{660}
   & \frac{1649}{660} & \frac{329}{660} & \frac{3139}{660} &
   \frac{1819}{660} & \frac{499}{660} & -\frac{1157}{660} &
   -\frac{1157}{660} & -\frac{1157}{660} & -\frac{329}{220} &
   -\frac{329}{220} & -\frac{329}{220} & \frac{1153}{220} &
   \frac{493}{220} & \frac{53}{220} \\
 \text{7b} & \frac{277}{130} & -\frac{147}{130} & \frac{3671}{780}
   & \frac{2111}{780} & \frac{551}{780} & \frac{3841}{780} &
   \frac{2281}{780} & \frac{721}{780} & -\frac{1043}{780} &
   -\frac{1043}{780} & -\frac{1043}{780} & -\frac{291}{260} &
   -\frac{291}{260} & -\frac{291}{260} & \frac{1427}{260} &
   \frac{647}{260} & \frac{127}{260} \\
   %
 \text{8a} & \frac{43}{22} & -\frac{21}{22} & \frac{607}{132} &
   \frac{343}{132} & \frac{79}{132} & \frac{641}{132} &
   \frac{377}{132} & \frac{113}{132} & -\frac{271}{132} &
   -\frac{271}{132} & -\frac{271}{132} & -\frac{79}{44} &
   -\frac{79}{44} & -\frac{79}{44} & \frac{235}{44} &
   \frac{103}{44} & \frac{15}{44} \\
 \text{8b} & \frac{303}{130} & -\frac{173}{130} & \frac{3749}{780}
   & \frac{2189}{780} & \frac{629}{780} & \frac{3919}{780} &
   \frac{2359}{780} & \frac{799}{780} & -\frac{1277}{780} &
   -\frac{1277}{780} & -\frac{1277}{780} & -\frac{369}{260} &
   -\frac{369}{260} & -\frac{369}{260} & \frac{1453}{260} &
   \frac{673}{260} & \frac{153}{260} \\
 \text{8c} & \frac{407}{150} & -\frac{257}{150} & \frac{4511}{900}
   & \frac{2711}{900} & \frac{911}{900} & \frac{4681}{900} &
   \frac{2881}{900} & \frac{1081}{900} & -\frac{1103}{900} &
   -\frac{1103}{900} & -\frac{1103}{900} & -\frac{311}{300} &
   -\frac{311}{300} & -\frac{311}{300} & \frac{1747}{300} &
   \frac{847}{300} & \frac{247}{300} \\
 \text{8d} & \frac{31}{10} & -\frac{21}{10} & \frac{313}{60} &
   \frac{193}{60} & \frac{73}{60} & \frac{323}{60} &
   \frac{203}{60} & \frac{83}{60} & -\frac{49}{60} &
   -\frac{49}{60} & -\frac{49}{60} & -\frac{13}{20} &
   -\frac{13}{20} & -\frac{13}{20} & \frac{121}{20} &
   \frac{61}{20} & \frac{21}{20} \\
   %
\text{9a} & \frac{237}{110} & -\frac{127}{110} &
   \frac{3101}{660} & \frac{1781}{660} & \frac{461}{660} &
   \frac{3271}{660} & \frac{1951}{660} & \frac{631}{660} &
   -\frac{1553}{660} & -\frac{1553}{660} & -\frac{1553}{660} &
   -\frac{461}{220} & -\frac{461}{220} & -\frac{461}{220} &
   \frac{1197}{220} & \frac{537}{220} & \frac{97}{220} \\
 \text{9b} & \frac{329}{130} & -\frac{199}{130} &
   \frac{3827}{780} & \frac{2267}{780} & \frac{707}{780} &
   \frac{3997}{780} & \frac{2437}{780} & \frac{877}{780} &
   -\frac{1511}{780} & -\frac{1511}{780} & -\frac{1511}{780} &
   -\frac{447}{260} & -\frac{447}{260} & -\frac{447}{260} &
   \frac{1479}{260} & \frac{699}{260} & \frac{179}{260} \\
 \text{9c} & \frac{437}{150} & -\frac{287}{150} &
   \frac{4601}{900} & \frac{2801}{900} & \frac{1001}{900} &
   \frac{4771}{900} & \frac{2971}{900} & \frac{1171}{900} &
   -\frac{1373}{900} & -\frac{1373}{900} & -\frac{1373}{900} &
   -\frac{401}{300} & -\frac{401}{300} & -\frac{401}{300} &
   \frac{1777}{300} & \frac{877}{300} & \frac{277}{300} \\
 \text{9d} & \frac{33}{10} & -\frac{23}{10} & \frac{319}{60} &
   \frac{199}{60} & \frac{79}{60} & \frac{329}{60} &
   \frac{209}{60} & \frac{89}{60} & -\frac{67}{60} &
   -\frac{67}{60} & -\frac{67}{60} & -\frac{19}{20} &
   -\frac{19}{20} & -\frac{19}{20} & \frac{123}{20} &
   \frac{63}{20} & \frac{23}{20} \\
   %
 \text{10a} & \frac{467}{150} & -\frac{317}{150} &
   \frac{4691}{900} & \frac{2891}{900} & \frac{1091}{900} &
   \frac{4861}{900} & \frac{3061}{900} & \frac{1261}{900} &
   -\frac{1643}{900} & -\frac{1643}{900} & -\frac{1643}{900} &
   -\frac{491}{300} & -\frac{491}{300} & -\frac{491}{300} &
   \frac{1807}{300} & \frac{907}{300} & \frac{307}{300} \\
 \text{10b} & \frac{7}{2} & -\frac{5}{2} & \frac{65}{12} &
   \frac{41}{12} & \frac{17}{12} & \frac{67}{12} & \frac{43}{12} &
   \frac{19}{12} & -\frac{17}{12} & -\frac{17}{12} &
   -\frac{17}{12} & -\frac{5}{4} & -\frac{5}{4} & -\frac{5}{4} &
   \frac{25}{4} & \frac{13}{4} & \frac{5}{4} \\
   %
 \text{11} & \frac{21}{11} & \frac{1}{11} & \frac{565}{132} &
   \frac{301}{132} & \frac{37}{132} & \frac{545}{132} &
   \frac{281}{132} & \frac{17}{132} & -\frac{91}{132} &
   -\frac{91}{132} & -\frac{91}{132} & -\frac{37}{44} &
   -\frac{37}{44} & -\frac{81}{44} & \frac{195}{44} &
   \frac{107}{44} & \frac{63}{44} \\
      %
 \text{12a} & \frac{23}{11} & -\frac{1}{11} & \frac{193}{44} &
   \frac{105}{44} & \frac{17}{44} & \frac{185}{44} & \frac{97}{44}
   & \frac{9}{44} & -\frac{43}{44} & -\frac{43}{44} &
   -\frac{43}{44} & -\frac{51}{44} & -\frac{51}{44} &
   -\frac{95}{44} & \frac{201}{44} & \frac{113}{44} &
   \frac{69}{44} \\
 \text{12b} & \frac{163}{65} & -\frac{33}{65} & \frac{1191}{260} &
   \frac{671}{260} & \frac{151}{260} & \frac{1151}{260} &
   \frac{631}{260} & \frac{111}{260} & -\frac{153}{260} &
   -\frac{153}{260} & -\frac{153}{260} & -\frac{193}{260} &
   -\frac{193}{260} & -\frac{453}{260} & \frac{1231}{260} &
   \frac{711}{260} & \frac{451}{260} \\
 \text{12c} & \frac{73}{25} & -\frac{23}{25} & \frac{1433}{300} &
   \frac{833}{300} & \frac{233}{300} & \frac{1393}{300} &
   \frac{793}{300} & \frac{193}{300} & -\frac{59}{300} &
   -\frac{59}{300} & -\frac{59}{300} & -\frac{33}{100} &
   -\frac{33}{100} & -\frac{133}{100} & \frac{491}{100} &
   \frac{291}{100} & \frac{191}{100} \\
   \bottomrule
\end{tabular}
\end{center}
\caption{Charge assignments of the fields under the $\text{U}(1)_R$ symmetry given the parameter values shown in Table~\ref{tab:collect-final} for the scenario where $y=1$.}
\label{tab:chargetabley1}
\end{table}

\clearpage

\bibliographystyle{JHEP}
\bibliography{jabrefbib.bib}

\providecommand{\href}[2]{#2}\begingroup\raggedright\begin{thebibliography}{10}

\bibitem{Aad:2012tfa}
{\bf ATLAS Collaboration} Collaboration, G.~Aad et~al., {\it {Observation of a
  new particle in the search for the Standard Model Higgs boson with the ATLAS
  detector at the LHC}},  {\em Phys.Lett.} {\bf B716} (2012) 1,
  [\href{http://xxx.lanl.gov/abs/1207.7214}{{\tt arXiv:1207.7214}}].

\bibitem{Chatrchyan:2012ufa}
{\bf CMS Collaboration} Collaboration, S.~Chatrchyan et~al., {\it {Observation
  of a new boson at a mass of 125~GeV with the CMS experiment at the LHC}},
  {\em Phys.Lett.} {\bf B716} (2012) 30,
  [\href{http://xxx.lanl.gov/abs/1207.7235}{{\tt arXiv:1207.7235}}].

\bibitem{Froggatt:1978nt}
C.~Froggatt and H.~B. Nielsen, {\it {Hierarchy of quark masses, Cabibbo angles
  and CP violation}},  {\em Nucl.Phys.} {\bf B147} (1979) 277.

\bibitem{Dreiner:2003yr}
H.~K. Dreiner, H.~Murayama, and M.~Thormeier, {\it {Anomalous flavour $\text
  U(1)_X$ for everything}},  {\em Nucl.Phys.} {\bf B729} (2005) 278,
  [\href{http://xxx.lanl.gov/abs/hep-ph/0312012}{{\tt hep-ph/0312012}}].

\bibitem{Dreiner:2006xw}
H.~K. Dreiner, C.~Luhn, H.~Murayama, and M.~Thormeier, {\it {Baryon triality
  and neutrino masses from an anomalous flavour $\text U(1)$}},  {\em
  Nucl.Phys.} {\bf B774} (2007) 127,
  [\href{http://xxx.lanl.gov/abs/hep-ph/0610026}{{\tt hep-ph/0610026}}].

\bibitem{Dreiner:2007vp}
H.~K. Dreiner, C.~Luhn, H.~Murayama, and M.~Thormeier, {\it {Proton hexality
  from an anomalous flavour $\text U(1)$ and neutrino masses: linking to the
  string scale}},  {\em Nucl.Phys.} {\bf B795} (2008) 172,
  [\href{http://xxx.lanl.gov/abs/0708.0989}{{\tt arXiv:0708.0989}}].

\bibitem{Lee:2010gv}
H.~M. Lee, S.~Raby, M.~Ratz, G.~G. Ross, R.~Schieren, et~al., {\it {A unique
  $\mathbb Z_4^R$ symmetry for the MSSM}},  {\em Phys.Lett.} {\bf B694} (2011)
  491, [\href{http://xxx.lanl.gov/abs/1009.0905}{{\tt arXiv:1009.0905}}].

\bibitem{Lee:2011dya}
H.~M. Lee, S.~Raby, M.~Ratz, G.~G. Ross, R.~Schieren, et~al., {\it {Discrete
  $R$-symmetries for the MSSM and its singlet extensions}},  {\em Nucl.Phys.}
  {\bf B850} (2011) 1, [\href{http://xxx.lanl.gov/abs/1102.3595}{{\tt
  arXiv:1102.3595}}].

\bibitem{Kurosawa:2001iq}
K.~Kurosawa, N.~Maru, and T.~Yanagida, {\it {Nonanomalous $R$-symmetry in
  supersymmetric unified theories of quarks and leptons}},  {\em Phys.Lett.}
  {\bf B512} (2001) 203, [\href{http://xxx.lanl.gov/abs/hep-ph/0105136}{{\tt
  hep-ph/0105136}}].

\bibitem{Babu:2002tx}
K.~Babu, I.~Gogoladze, and K.~Wang, {\it {Natural $R$-parity, $\mu$-term, and
  fermion mass hierarchy from discrete gauge symmetries}},  {\em Nucl.Phys.}
  {\bf B660} (2003) 322, [\href{http://xxx.lanl.gov/abs/hep-ph/0212245}{{\tt
  hep-ph/0212245}}].

\bibitem{Green:1984sg}
M.~B. Green and J.~H. Schwarz, {\it {Anomaly cancellation in supersymmetric
  $d=10$ gauge theory and superstring theory}},  {\em Phys.Lett.} {\bf B149}
  (1984) 117.

\bibitem{Kim:1983dt}
J.~E. Kim and H.~P. Nilles, {\it {The $\mu$-problem and the strong CP
  problem}},  {\em Phys.Lett.} {\bf B138} (1984) 150.

\bibitem{Giudice:1988yz}
G.~Giudice and A.~Masiero, {\it {A natural solution to the $\mu$-problem in
  supergravity theories}},  {\em Phys.Lett.} {\bf B206} (1988) 480.

\bibitem{Kim:1994eu}
J.~E. Kim and H.~P. Nilles, {\it {Symmetry principles toward solutions of the
  $\mu$-problem}},  {\em Mod.Phys.Lett.} {\bf A9} (1994) 3575,
  [\href{http://xxx.lanl.gov/abs/hep-ph/9406296}{{\tt hep-ph/9406296}}].

\bibitem{Dine:1986zy}
M.~Dine, N.~Seiberg, X.~Wen, and E.~Witten, {\it {Nonperturbative effects on
  the string world sheet}},  {\em Nucl.Phys.} {\bf B278} (1986) 769.

\bibitem{Dine:1987bq}
M.~Dine, N.~Seiberg, X.~Wen, and E.~Witten, {\it {Nonperturbative effects on
  the string world sheet 2}},  {\em Nucl.Phys.} {\bf B289} (1987) 319.

\bibitem{Atick:1987gy}
J.~J. Atick, L.~J. Dixon, and A.~Sen, {\it {String calculation of
  Fayet-Iliopoulos $D$-terms in arbitrary supersymmetric compactifications}},
  {\em Nucl.Phys.} {\bf B292} (1987) 109.

\bibitem{Dine:1987gj}
M.~Dine, I.~Ichinose, and N.~Seiberg, {\it {$F$-terms and $D$-terms in string
  theory}},  {\em Nucl.Phys.} {\bf B293} (1987) 253.

\bibitem{Freedman:1976uk}
D.~Z. Freedman, {\it {Supergravity with axial gauge invariance}},  {\em
  Phys.Rev.} {\bf D15} (1977) 1173.

\bibitem{Chamseddine:1995gb}
A.~H. Chamseddine and H.~K. Dreiner, {\it {Anomaly-free gauged $R$-symmetry in
  local supersymmetry}},  {\em Nucl.Phys.} {\bf B458} (1996) 65,
  [\href{http://xxx.lanl.gov/abs/hep-ph/9504337}{{\tt hep-ph/9504337}}].

\bibitem{Castano:1995ci}
D.~Casta\~no, D.~Freedman, and C.~Manuel, {\it {Consequences of supergravity
  with gauged $\text U(1)_R$ symmetry}},  {\em Nucl.Phys.} {\bf B461} (1996)
  50, [\href{http://xxx.lanl.gov/abs/hep-ph/9507397}{{\tt hep-ph/9507397}}].

\bibitem{CDP}
E.~H. Wichmann and J.~H. Crichton, {\it {Cluster decomposition properties of
  the S-matrix}},  {\em Phys.Rev.} {\bf 132} (1963) 2788.

\bibitem{Weinberg:1996kw}
S.~Weinberg, {\it {What is quantum field theory, and what did we think it
  is?}},  \href{http://xxx.lanl.gov/abs/hep-th/9702027}{{\tt hep-th/9702027}}.

\bibitem{Wolfenstein:1983yz}
L.~Wolfenstein, {\it {Parametrisation of the Kobayashi-Maskawa matrix}},  {\em
  Phys.Rev.Lett.} {\bf 51} (1983) 1945.

\bibitem{Ramond:1993kv}
P.~Ramond, R.~Roberts, and G.~G. Ross, {\it {Stitching the Yukawa quilt}},
  {\em Nucl.Phys.} {\bf B406} (1993) 19,
  [\href{http://xxx.lanl.gov/abs/hep-ph/9303320}{{\tt hep-ph/9303320}}].

\bibitem{Binetruy:1994ru}
P.~Bin\'etruy and P.~Ramond, {\it {Yukawa textures and anomalies}},  {\em
  Phys.Lett.} {\bf B350} (1995) 49,
  [\href{http://xxx.lanl.gov/abs/hep-ph/9412385}{{\tt hep-ph/9412385}}].

\bibitem{Dudas:1995yu}
E.~Dudas, S.~Pokorski, and C.~A. Savoy, {\it {Yukawa matrices from a
  spontaneously broken Abelian symmetry}},  {\em Phys.Lett.} {\bf B356} (1995)
  45, [\href{http://xxx.lanl.gov/abs/hep-ph/9504292}{{\tt hep-ph/9504292}}].

\bibitem{Nir:1995bu}
Y.~Nir, {\it {Gauge unification, Yukawa hierarchy and the $\mu$-problem}},
  {\em Phys.Lett.} {\bf B354} (1995) 107,
  [\href{http://xxx.lanl.gov/abs/hep-ph/9504312}{{\tt hep-ph/9504312}}].

\bibitem{Binetruy:1996xk}
P.~Bin\'etruy, S.~Lavignac, and P.~Ramond, {\it {Yukawa textures with an
  anomalous horizontal Abelian symmetry}},  {\em Nucl.Phys.} {\bf B477} (1996)
  353, [\href{http://xxx.lanl.gov/abs/hep-ph/9601243}{{\tt hep-ph/9601243}}].

\bibitem{Irges:1998ax}
N.~Irges, S.~Lavignac, and P.~Ramond, {\it {Predictions from an anomalous
  $\text U(1)$ model of Yukawa hierarchies}},  {\em Phys.Rev.} {\bf D58} (1998)
  035003, [\href{http://xxx.lanl.gov/abs/hep-ph/9802334}{{\tt
  hep-ph/9802334}}].

\bibitem{Jack:2002pn}
I.~Jack, D.~Jones, and R.~Wild, {\it {Fayet-Iliopoulos $D$-terms, neutrino
  masses and anomaly mediated supersymmetry breaking}},  {\em Phys.Lett.} {\bf
  B535} (2002) 193, [\href{http://xxx.lanl.gov/abs/hep-ph/0202101}{{\tt
  hep-ph/0202101}}].

\bibitem{Dreiner:2003hw}
H.~K. Dreiner and M.~Thormeier, {\it {Supersymmetric Froggatt-Nielsen models
  with baryon and lepton number violation}},  {\em Phys.Rev.} {\bf D69} (2004)
  053002, [\href{http://xxx.lanl.gov/abs/hep-ph/0305270}{{\tt
  hep-ph/0305270}}].

\bibitem{Dreiner:2008tw}
H.~K. Dreiner, H.~E. Haber, and S.~P. Martin, {\it {Two-component spinor
  techniques and Feynman rules for quantum field theory and supersymmetry}},
  {\em Phys.Rept.} {\bf 494} (2010) 1,
  [\href{http://xxx.lanl.gov/abs/0812.1594}{{\tt arXiv:0812.1594}}].

\bibitem{Jack:2003pb}
I.~Jack, D.~Jones, and R.~Wild, {\it {Yukawa textures and the $\mu$-term}},
  {\em Phys.Lett.} {\bf B580} (2004) 72,
  [\href{http://xxx.lanl.gov/abs/hep-ph/0309165}{{\tt hep-ph/0309165}}].

\bibitem{Leurer:1992wg}
M.~Leurer, Y.~Nir, and N.~Seiberg, {\it {Mass matrix models}},  {\em
  Nucl.Phys.} {\bf B398} (1993) 319,
  [\href{http://xxx.lanl.gov/abs/hep-ph/9212278}{{\tt hep-ph/9212278}}].

\bibitem{Brummer:2010fr}
F.~Br{\"u}mmer, R.~Kappl, M.~Ratz, and K.~Schmidt-Hoberg, {\it {Approximate
  $R$-symmetries and the $\mu$-term}},  {\em JHEP} {\bf 1004} (2010) 006,
  [\href{http://xxx.lanl.gov/abs/1003.0084}{{\tt arXiv:1003.0084}}].

\bibitem{Chen:2012jg}
M.-C. Chen, M.~Ratz, C.~Staudt, and P.~K. Vaudrevange, {\it {The $\mu$-term and
  neutrino masses}},  {\em Nucl.Phys.} {\bf B866} (2013) 157,
  [\href{http://xxx.lanl.gov/abs/1206.5375}{{\tt arXiv:1206.5375}}].

\bibitem{Ferrara:1982qs}
S.~Ferrara, L.~Girardello, and H.~P. Nilles, {\it {Breakdown of local
  supersymmetry through gauge fermion condensates}},  {\em Phys.Lett.} {\bf
  B125} (1983) 457.

\bibitem{Derendinger:1985kk}
J.~Derendinger, L.~E. Ib\'a\~nez, and H.~P. Nilles, {\it {On the low-energy $d
  = 4$, $N=1$ supergravity theory extracted from the $d = 10$, $N=1$
  superstring}},  {\em Phys.Lett.} {\bf B155} (1985) 65.

\bibitem{Dine:1985rz}
M.~Dine, R.~Rohm, N.~Seiberg, and E.~Witten, {\it {Gluino condensation in
  superstring models}},  {\em Phys.Lett.} {\bf B156} (1985) 55.

\bibitem{Ibanez:1992fy}
L.~E. Ib\'a\~nez, {\it {Computing the weak mixing angle from anomaly
  cancellation}},  {\em Phys.Lett.} {\bf B303} (1993) 55,
  [\href{http://xxx.lanl.gov/abs/hep-ph/9205234}{{\tt hep-ph/9205234}}].

\bibitem{Dreiner:2012ae}
H.~K. Dreiner, M.~Hanussek, and C.~Luhn, {\it {What is the discrete gauge
  symmetry of the $R$-parity violating MSSM?}},  {\em Phys.Rev.} {\bf D86}
  (2012) 055012, [\href{http://xxx.lanl.gov/abs/1206.6305}{{\tt
  arXiv:1206.6305}}].

\bibitem{Ginsparg:1987ee}
P.~H. Ginsparg, {\it {Gauge and gravitational couplings in four-dimensional
  string theories}},  {\em Phys.Lett.} {\bf B197} (1987) 139.

\bibitem{Banks:1991xj}
T.~Banks and M.~Dine, {\it {Note on discrete gauge anomalies}},  {\em
  Phys.Rev.} {\bf D45} (1992) 1424,
  [\href{http://xxx.lanl.gov/abs/hep-th/9109045}{{\tt hep-th/9109045}}].

\bibitem{Minkowski:1977sc}
P.~Minkowski, {\it {$\mu \rightarrow e \gamma$ at a rate of one out of 1
  billion muon decays?}},  {\em Phys.Lett.} {\bf B67} (1977) 421.

\bibitem{ramond-seesaw}
M.~Gell-Mann, P.~Ramond, and R.~Slansky, {\em {\rm{in Sanibel Talk,
  CALT-68-709, hep-ph/9809459, and in {\it Supergravity}, North-Holland,
  Amsterdam (1979)}}}.

\bibitem{yanagida-seesaw}
T.~Yanagida, {\em {Proceedings of the Workshop on Unified Theory and Baryon
  Number of the Universe, \rm{KEK, Japan (1979)}}}.

\bibitem{Mohapatra:1979ia}
R.~N. Mohapatra and G.~Senjanovic, {\it {Neutrino mass and spontaneous parity
  violation}},  {\em Phys.Rev.Lett.} {\bf 44} (1980) 912.

\bibitem{Weinberg:1979sa}
S.~Weinberg, {\it {Baryon and lepton nonconserving processes}},  {\em
  Phys.Rev.Lett.} {\bf 43} (1979) 1566.

\bibitem{Tortola:2012te}
D.~Forero, M.~Tortola, and J.~Valle, {\it {Global status of neutrino
  oscillation parameters after Neutrino 2012}},  {\em Phys.Rev.} {\bf D86}
  (2012) 073012, [\href{http://xxx.lanl.gov/abs/1205.4018}{{\tt
  arXiv:1205.4018}}].

\bibitem{Fogli:2012ua}
G.~Fogli, E.~Lisi, A.~Marrone, D.~Montanino, A.~Palazzo, et~al., {\it {Global
  analysis of neutrino masses, mixings and phases: entering the era of leptonic
  CP violation searches}},  {\em Phys.Rev.} {\bf D86} (2012) 013012,
  [\href{http://xxx.lanl.gov/abs/1205.5254}{{\tt arXiv:1205.5254}}].

\bibitem{GonzalezGarcia:2012sz}
M.~Gonzalez-Garcia, M.~Maltoni, J.~Salvado, and T.~Schwetz, {\it {Global fit to
  three neutrino mixing: critical look at present precision}},  {\em JHEP} {\bf
  1212} (2012) 123, [\href{http://xxx.lanl.gov/abs/1209.3023}{{\tt
  arXiv:1209.3023}}].

\bibitem{Luhn:2007yr}
C.~Luhn, S.~Nasri, and P.~Ramond, {\it {Simple finite non-Abelian flavour
  groups}},  {\em J.Math.Phys.} {\bf 48} (2007) 123519,
  [\href{http://xxx.lanl.gov/abs/0709.1447}{{\tt arXiv:0709.1447}}].

\bibitem{Ludl:2009ft}
P.~O. Ludl, {\it {Systematic analysis of finite family symmetry groups and
  their application to the lepton sector}},
  \href{http://xxx.lanl.gov/abs/0907.5587}{{\tt arXiv:0907.5587}}.

\bibitem{Altarelli:2010gt}
G.~Altarelli and F.~Feruglio, {\it {Discrete flavour symmetries and models of
  neutrino mixing}},  {\em Rev.Mod.Phys.} {\bf 82} (2010) 2701,
  [\href{http://xxx.lanl.gov/abs/1002.0211}{{\tt arXiv:1002.0211}}].

\bibitem{Ishimori:2010au}
H.~Ishimori, T.~Kobayashi, H.~Ohki, Y.~Shimizu, H.~Okada, et~al., {\it
  {Non-Abelian discrete symmetries in particle physics}},  {\em
  Prog.Theor.Phys.Suppl.} {\bf 183} (2010) 1,
  [\href{http://xxx.lanl.gov/abs/1003.3552}{{\tt arXiv:1003.3552}}].

\bibitem{Grimus:2011fk}
W.~Grimus and P.~O. Ludl, {\it {Finite flavour groups of fermions}},  {\em
  J.Phys.} {\bf A45} (2012) 233001,
  [\href{http://xxx.lanl.gov/abs/1110.6376}{{\tt arXiv:1110.6376}}].

\bibitem{King:2013eh}
S.~F. King and C.~Luhn, {\it {Neutrino mass and mixing with discrete
  symmetry}},  {\em Rept.Prog.Phys.} {\bf 76} (2013) 056201,
  [\href{http://xxx.lanl.gov/abs/1301.1340}{{\tt arXiv:1301.1340}}].

\bibitem{GonzalezGarcia:2007ib}
M.~Gonzalez-Garcia and M.~Maltoni, {\it {Phenomenology with massive
  neutrinos}},  {\em Phys.Rept.} {\bf 460} (2008) 1,
  [\href{http://xxx.lanl.gov/abs/0704.1800}{{\tt arXiv:0704.1800}}].

\bibitem{Fukugita:1986hr}
M.~Fukugita and T.~Yanagida, {\it {Baryogenesis without grand unification}},
  {\em Phys.Lett.} {\bf B174} (1986) 45.

\bibitem{Davidson:2002qv}
S.~Davidson and A.~Ibarra, {\it {A lower bound on the right-handed neutrino
  mass from leptogenesis}},  {\em Phys.Lett.} {\bf B535} (2002) 25,
  [\href{http://xxx.lanl.gov/abs/hep-ph/0202239}{{\tt hep-ph/0202239}}].

\bibitem{Buchmuller:2002rq}
W.~Buchm{\"u}ller, P.~Di~Bari, and M.~Pl{\"u}macher, {\it {Cosmic microwave
  background, matter-antimatter asymmetry and neutrino masses}},  {\em
  Nucl.Phys.} {\bf B643} (2002) 367,
  [\href{http://xxx.lanl.gov/abs/hep-ph/0205349}{{\tt hep-ph/0205349}}].

\bibitem{Buchmuller:2004nz}
W.~Buchm{\"u}ller, P.~Di~Bari, and M.~Pl{\"u}macher, {\it {Leptogenesis for
  pedestrians}},  {\em Annals Phys.} {\bf 315} (2005) 305,
  [\href{http://xxx.lanl.gov/abs/hep-ph/0401240}{{\tt hep-ph/0401240}}].

\end{thebibliography}\endgroup

\end{document}